\documentclass[twocolumn,trackchanges]{aastex62}
\usepackage{amsmath}

\newcommand\kms{km s$^{-1}$}
\newcommand\masyr{mas yr$^{-1}$}

\usepackage{listings}

\begin{document}
\title{Untangling the Galaxy I: Local Structure and Star Formation History of the Milky Way}

\author[0000-0002-5365-1267]{Marina Kounkel}
\affil{Department of Physics and Astronomy, Western Washington University, 516 High St, Bellingham, WA 98225}
\author[0000-0001-6914-7797]{Kevin Covey}
\affil{Department of Physics and Astronomy, Western Washington University, 516 High St, Bellingham, WA 98225}
\email{marina.kounkel@wwu.edu}

\begin{abstract}
\textit{Gaia} DR2 provides unprecedented precision in measurements of the distance and kinematics of stars in the solar neighborhood. Through applying unsupervised machine learning on DR2's 5-dimensional dataset (3d position + 2d velocity), we identify a number of clusters, associations, and co-moving groups within 1 kpc and $|b|<30^\circ$ (many of which have not been previously known). We estimate their ages with the precision of $\sim$0.15 dex. Many of these groups appear to be filamentary or string-like, oriented in parallel to the Galactic plane, and some span hundreds of pc in length. Most of these string lack a central cluster, indicating that their filamentary structure is primordial, rather than the result of tidal stripping or dynamical processing. The youngest strings ($<$100 Myr) are orthogonal to the Local Arm. The older ones appear to be remnants of several other arm-like structures that cannot be presently traced by dust and gas. The velocity dispersion measured from the ensemble of groups and strings increase with age, suggesting a timescale for dynamical heating of $\sim$300 Myr. This timescale is also consistent with the age at which the population of strings begins to decline, while the population in more compact groups continues to increase, suggesting that dynamical processes are disrupting the weakly bound string populations, leaving only individual clusters to be identified at the oldest ages. These data shed a new light on the local galactic structure and a large scale cloud collapse.
\end{abstract}

\keywords{Milky Way dynamics (1051), Galaxy structure (622), Stellar kinematics (1608), Star clusters (1567), Stellar associations (1582), Stellar ages (1581)}

\section{Introduction}

It is typically recognized that stars tend to form in clustered environments \citep{lada2003}, although the sizes and longevity of these clusters do vary significantly. In the solar neighborhood, however, there are a number of examples of more diffuse modes of star formation as well. The closest young OB association, Sco OB2, has only a few dense subclusters; most of the young stars belong to various diffuse populations \citep{damiani2019}. Another nearby massive star-forming complex, Orion, has a number of massive clusters (including the Orion Nebula), but it has a significant distributed component as well \citep{kounkel2018a}. The Taurus Molecular Clouds are notable for not having any major clusters \citep{luhman2018}. 

After star forming molecular clouds dissipate, stars outside of the most massive clusters are expected to disperse throughout the galaxy. The timescales of this process are not yet fully constrained. However, because a newly-born population of stars would form from the same molecular cloud, it would have very similar kinematics (to within a few \kms), and it may take several tens if not hundreds of Myr for the population to fully lose coherence. 

While there is a long-standing history of studying extended populations in young ($<10$ Myr) star forming regions, studies of evolved structure is primarily limited to dense clusters. Some evolved non-clustered moving groups have been previously identified, but they were typically found within 100 pc. They include regions such as TW Hya, Tuc–Hor, AB Dor, Ursa Major, and several others \citep[e.g.,][]{bannister2005,bell2015,faherty2018,lee2018}. Few such populations are known at larger distances, as until recently, no sufficiently precise data on stellar kinematics were available.

This has recently changed with the release of \textit{Gaia} DR2 \citep{gaia-collaboration2018}, which provides unprecedented precision and sensitivity in measurements of stellar parallaxes and proper motions for 1.3 billion stars. On the bright end ($G<15$) it can achieve precision to better than 40 $\mu$as in parallax and 60 $\mu$as yr$^{-1}$ in proper motions. On the faint end ($G>20$) the typical uncertainties are 0.7 mas in parallax and 1.2 \masyr\ in proper motions. In addition, \textit{Gaia} provides radial velocity (RV) measurements for 7.2 million stars \citep{cropper2018} with $R\sim11,500$ and a typical uncertainty of $\sim$2 \kms. Although neither the quantity nor the precision of the RV portion of the survey compares with the astrometric portion, it is (currently) the only large spectroscopic program that provides observations across the entire sky.

Using data from \textit{Gaia} DR2, a number of studies of both individual clusters and their bulk distributions have been conducted, and a number of new open clusters have been found \citep[e.g.,][]{cantat-gaudin2018a,castro-ginard2019}, showing that the current census of open clusters is still incomplete. Some studies focused on tracing the spiral arm structure and kinematics of the Milky Way \citep[e.g.,][]{dias2019,bobylev2019a} using known clusters. There have also been studies of the overall spatial density distribution of young stars in the solar neighborhood \citep{zari2018}, as well as detailed characterization of the structure and dynamics of individual star-forming regions \citep[e.g.,][]{kounkel2018a,damiani2019,cantat-gaudin2019}.

In this paper, we identify extended structures, estimate their ages, and analyze their distribution. In Section \ref{sec:data} we describe the data selection process and the clustering that was performed. In Section \ref{sec:ages} we discuss the estimation of ages in the identified groups. In Section \ref{sec:string} we describe the classification of groups that correspond to extended large-scale structures, their individual properties. We also focus on the distribution of these extended structures in the context of the Galaxy, and discuss the possible implications for the dynamical evolution of the Milky Way. Finally, in Section \ref{sec:concl} we offer our conclusions. This paper specifically focuses on the generation of the catalog of members of co-moving groups within 1 kpc and their characterization. Their analysis in the theoretical framework of the Galactic star formation, structure, and dynamics is deferred to subsequent works in the series.

\section{Data}\label{sec:data}

\begin{deluxetable}{cccc}
\tabletypesize{\scriptsize}
\tablewidth{0pt}
\tablecaption{Clustered sources\label{tab:data}}
\tablehead{
\colhead{Gaia DR2} &\colhead{$\alpha$} & \colhead{$\delta$} & \colhead{Theia}\\
\colhead{ID} &\colhead{(deg)} & \colhead{(deg)} & \colhead{Group ID}
}
\startdata
2172342682494385024 & 320.57220379 & 52.07733956 & 1 \\
2170224782576556672 & 315.65202897 & 52.20035062 & 1 \\
2168760576695182208 & 315.71659886 & 50.22889438 & 1
\enddata
\tablenotetext{}{Only a portion shown here. Full table with all deconvolved parameters is available in an electronic form.}
\end{deluxetable}

\begin{splitdeluxetable*}{cccccccccccBccccccccc}
\tabletypesize{\scriptsize}
\tablewidth{0pt}
\tablecaption{Identified structures and their average properties \label{tab:cluster}}
\tablehead{
\colhead{Theia} &\colhead{Common} & \colhead{Age} &\colhead{$A_V$} & \colhead{String?} & \colhead{Projected} & \colhead{Projected} & \colhead{String}& \colhead{$l$}& \colhead{$b$}& \colhead{$\pi$} & \colhead{$v_l$}& \colhead{$v_b$}& \colhead{$v_r$} & \colhead{$X$}& \colhead{$Y$}& \colhead{$Z$} & \colhead{$U$}& \colhead{$V$}& \colhead{$W$}\\
\colhead{Group ID} &\colhead{name} & \colhead{(dex)} & \colhead{(mag)} & \colhead{} & \colhead{width (pc)} & \colhead{height (pc)}& \colhead{length (pc)}& \colhead{(deg)}& \colhead{(deg)}& \colhead{(mas)}& \colhead{(\masyr)}& \colhead{(\masyr)}& \colhead{(\kms)}& \colhead{(pc)}& \colhead{(pc)}& \colhead{(pc)}& \colhead{(\kms)}& \colhead{(\kms)}& \colhead{(\kms)}
}
\startdata
1 & LDN 988e & 6.1 & 1.7 &  &  9.7 &  7.1 &   &   90.922 &   2.794 &  1.677 & -16.354 &  -3.154 &  -9.3 &   -9.6 &  595.5 &   29.1 &  17.2 &  -9.0 &  -3.7 \\
9 & IC 1396 & 6.6 & 1.2 & y & 47.2 &  8.9 & 130.4 &  102.828 &   4.276 &  1.104 & -29.085 &  -0.536 &   6.9 & -207.3 &  880.8 &   68.3 &  27.8 &  12.7 &   0.3 \\
16 &  & 6.9 & 1.5 &  &  4.2 &  4.8 &   0.0 &  319.605 &   1.527 &  1.089 &  -7.130 &   3.015 &  -0.6 &  698.9 & -594.7 &   24.5 &  -5.4 &  -5.3 &   3.2   \\
\enddata
\tablenotetext{}{Only a portion shown here. Full table with all deconvolved parameters is available in an electronic form.}
\end{splitdeluxetable*}

\begin{deluxetable*}{ccccccccccccc}
\tabletypesize{\scriptsize}
\tablewidth{0pt}
\tablecaption{Traces of the strings \label{tab:string}}
\tablehead{
\colhead{Theia} & \colhead{$l$}& \colhead{$b$}& \colhead{$\pi$} & \colhead{$v_l$}& \colhead{$v_b$}& \colhead{$v_r$} & \colhead{$X$}& \colhead{$Y$}& \colhead{$Z$} & \colhead{$U$}& \colhead{$V$}& \colhead{$W$}\\
\colhead{Group ID} & \colhead{(deg)}& \colhead{(deg)}& \colhead{(mas)}& \colhead{(\masyr)}& \colhead{(\masyr)}& \colhead{(\kms)}& \colhead{(pc)}& \colhead{(pc)}& \colhead{(pc)}& \colhead{(\kms)}& \colhead{(\kms)}& \colhead{(\kms)}
}
\startdata
9 &     98.5 &   3.873 &  1.102 &  -8.570 &  -0.339 &   8.4 & -133.9 &  895.8 &   61.3 &  36.0 &  14.1 &  -0.9 \\ 
9 &     99.5 &   3.771 &  1.097 &  -8.025 &  -0.403 &   8.4 & -150.1 &  897.2 &   60.0 &  33.6 &  14.4 &  -1.2 \\ 
9 &    100.5 &   3.955 &  1.097 &  -7.352 &  -0.300 &   9.1 & -165.7 &  894.1 &   62.9 &  30.2 &  15.1 &  -0.7 \\ 
\enddata
\tablenotetext{}{Only a portion shown here. Full table with all deconvolved parameters is available in an electronic form.}
\end{deluxetable*}

We used \textit{Gaia} DR2 data for the analysis. We focus on the Galactic midplane, where star formation persists today, and the extended solar neighborhood, where extended structures will have the largest angular size. We select sources within $|b|<30^\circ$ of the Galactic plane, and $\pi>1$ mas. We further limit our sample to sources with robust astrometric and photometric detections, requiring

\begin{itemize}
\item $\sigma_\pi<0.1$ mas or $\sigma_\pi/\pi<0.1$
\item 1.0857 / phot\_g\_mean\_flux\_over\_error $<$ 0.03
\item astrometric\_sigma5d\_max $<$ 0.3
\item visibility\_periods\_used $>$8
\item astrometric\_excess\_noise$<$1  or \\(astrometric\_excess\_noise$>$1 and \\astrometric\_excess\_noise\_sig$<$2)
\item ${v_{\alpha,\delta}^{lsr}}<60$ \kms
\end{itemize}

\noindent These cuts ensure a high quality dataset for our hierarchical clustering algorithm, which may perform poorly with data that are very uncertain. The resulting catalog consisted of 19.55 million stars, typically with $G<18$ mag. The $|b|<30^\circ$ cut excludes the space that is not expected to contain a significant number of clusters and co-moving groups. The ones that are present at higher galactic latitudes would also be more difficult to recover with the chosen algorithm due to the distortions in $l$ from $\cos b$, and having strong outliers may have a negative effect on the recovery of the remaining structures (this is the same motivation for imposing the proper motion limit). The $|b|<30^\circ$ cut-off was set to include everything up to and including the structures considered to be part of the Gould Belt.

We perform a clustering analysis on this dataset using Python implementation of HDBSCAN \citep[Hierarchical Density-Based Spatial Clustering of Applications with Noise,][]{hdbscan}. It is a hierarchical clustering algorithm adapted from the more commonly used DBSCAN \citep{dbscan}. DBSCAN identifies clusters as overdensities in a multi-dimensional space in which the number of sources exceeds the required minimum number of points within a neighborhood of a particular linking length $\epsilon$. HDBSCAN does not depend on $\epsilon$; instead it condenses the minimum spanning tree by pruning off the nodes that do not meet the minimum number of sources in a cluster, and reanalyzing the nodes that do. Depending on the chosen algorithm, it would then either find the most persistent structure (through the excess of mass method), or return clusters as the leaves of the tree (which results in somewhat more homogeneous clusters). In both cases it is more effective at finding structures of varying densities in a given dataset than DBSCAN.

The two main parameters that control HDBSCAN are the number of sources in a cluster, and the number of samples. The former is the parameter that rejects groupings that are too small; the latter sets the threshold of how conservative the algorithm is in its considerations of the background noise (even if the resulting noisy groupings do meet the minimum cluster size). By default, the sample size is set to the same value as the cluster size, but it is possibly to adjust them separately\footnote{\url{https://hdbscan.readthedocs.io/en/latest/index.html}}.

The clustering was performed on the 5-dimensional dataset: Galactic coordinates $l$ and $b$, parallax $\pi$, and proper motions. The conversion from the equatorial to the Galactic reference frame for the positions themselves is necessary, as most of the structure is located along the Galactic plane, and the $\cos\delta$ term would add non-linear distortions in $\alpha$ otherwise. However, in terms of the proper motions, the combinations of $v_\alpha,v_\delta$ and $v_l,v_b$ are the direct rotational transformation of each other, and thus they produce largerly comparable outputs. Because the range of scatter in $v_l$ is typically larger than in $v_b$, but $v_\alpha$ and $v_\delta$ tend to have similar ranges, the latter was chosen. Proper motions were converted to the local standard of rest (lsr) reference frame \citep[using constants from ][]{schonrich2010} to avoid distortions due to the line of sight from the solar motion, as well as converted to the physical units of \kms\ to avoid the distortion in the distance.

Various scaling factors were considered to normalize each of the 5 dimensions, but they did not have a strong effect on the resulting groups, and they were left in their native units (i.e., degrees, mas, and \kms). We also considered additional transformations, such as distance instead of parallax or the XYZ positions, but they were not considered optimal due to a greater degree of noise in the outputs. We required a minimum number of samples to be 25 sources, with the minimum number of stars per cluster to be 40 stars.

\begin{figure*}
\epsscale{1.1}
\plottwo{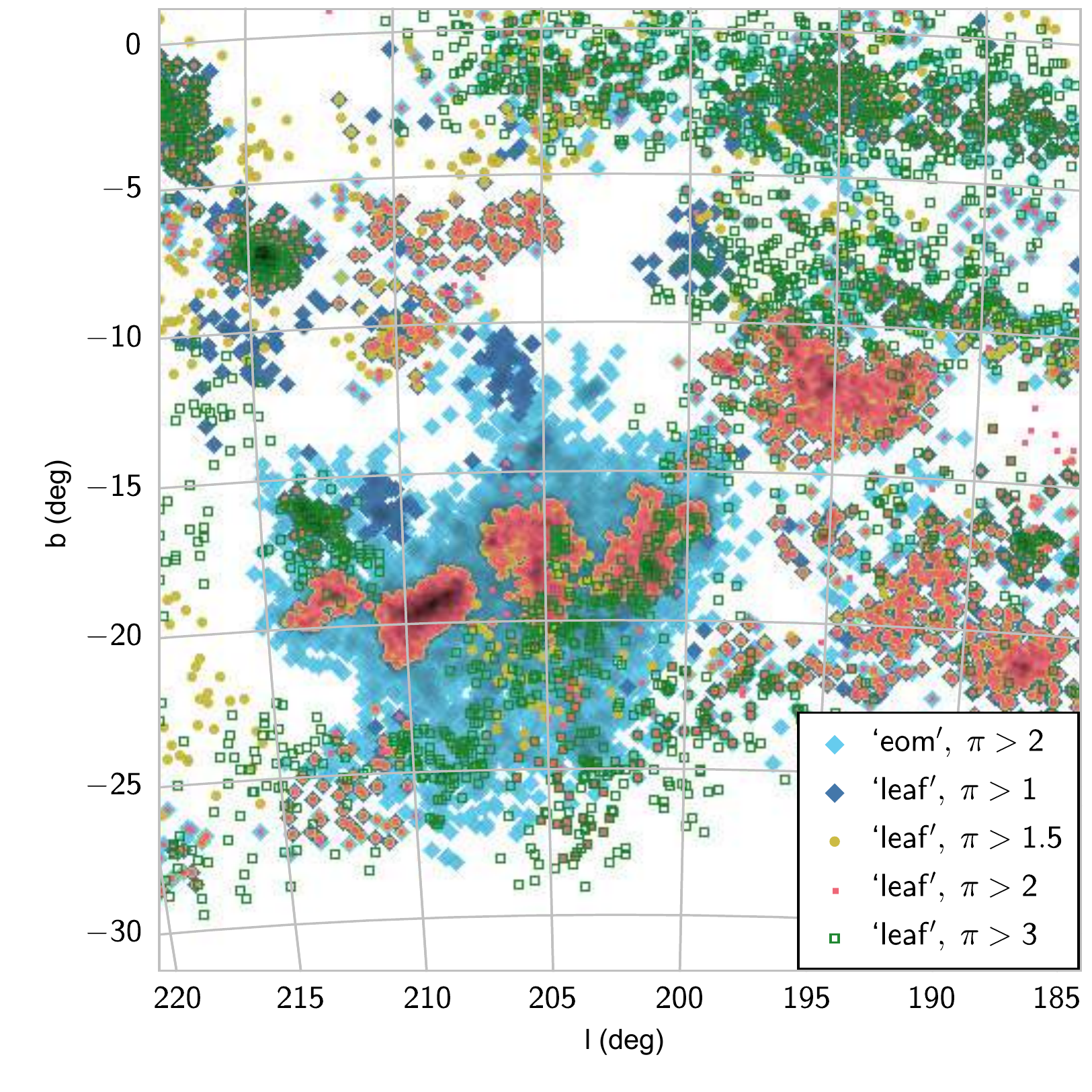}{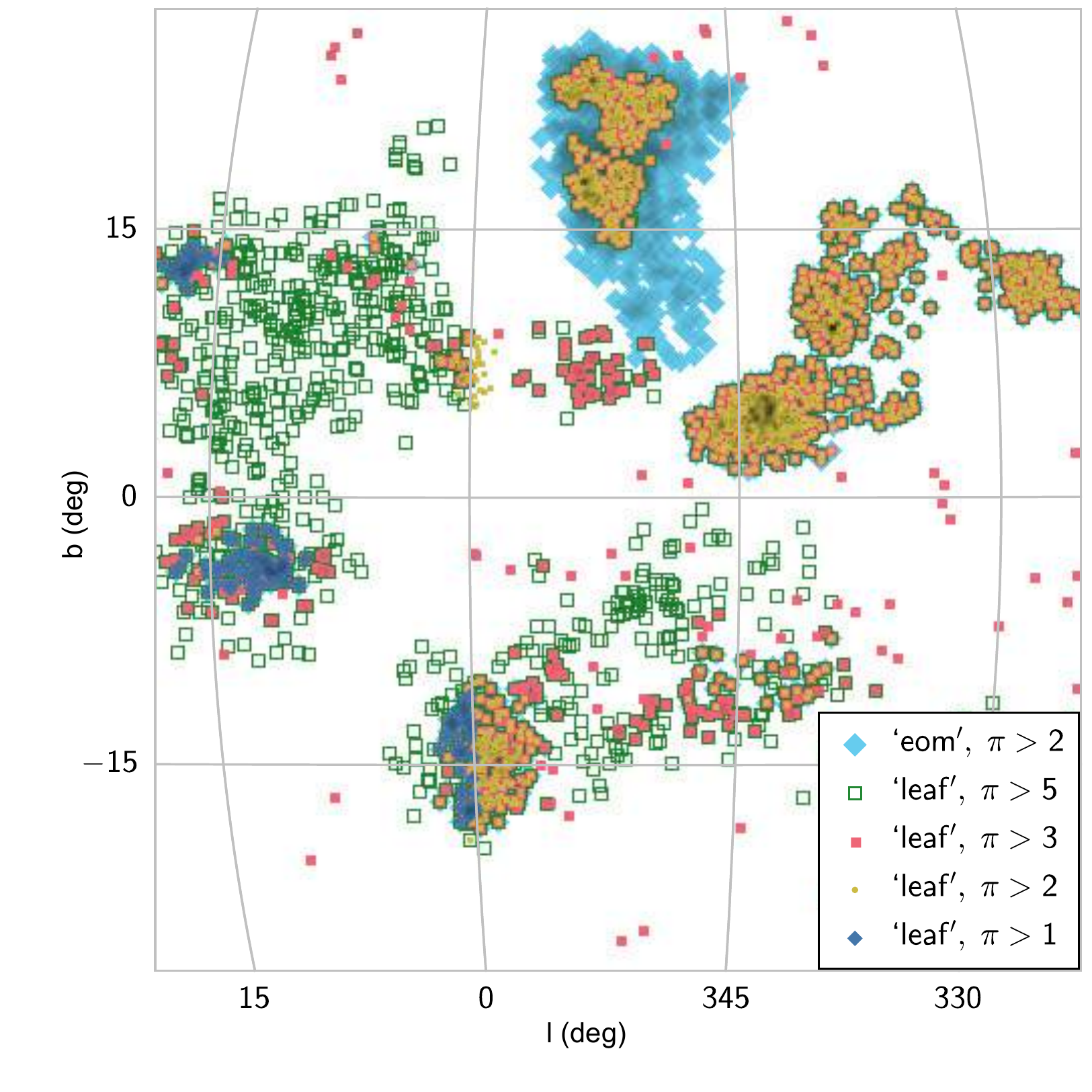}
\caption{Comparison of HDBSCAN outputs using different clustering methods and different cut-off parallaxes. Left: Orion, sources shown only up to $\pi=$2 mas. Right: Upper Sco and CrA, sources shown only up to $\pi=$5 mas. Both panels are shown in Galactic coordinates. Different symbols indicate different products of different clustering runs. The structures they trace vary depending on the cut-off parallax, including the persistence of various structures and their specific membership. Note the edge effects at $l=0^\circ$ in the runs shown in the right panel.
\label{fig:comp}}
\end{figure*}

While HDBSCAN is able to robustly recover structures of very different densities and sizes, the resulting distance matrix that is computed is strongly correlated with distance. The further a cluster is, the smaller it is going to be in the plane of the sky, the smaller the proper motions are going to be (and they will have more scatter due to uncertainties when transformed to the velocity space), and the smaller the parallaxes would be. Combined with the fact that there is more volume of space to encompass the structures that are more distant than those that are nearby, any clustering runs that include distant sources (i.e. have a smaller cut-off parallax) will be less attuned to the apparent densities of the nearest groups. The cores of the nearby massive clusters (e.g. Pleiades, Hyades) can be recovered in all the runs, regardless of the cut-off parallax, but the sources in their periphery, or some of the lower density structures have a significantly poorer recovery in clustering runs that extend to our full distance limit (1 kpc). We therefore performed several different runs with different cut-offs in parallax --- 10, 9, 8, 7, 6, 5, 4, 3, 2, 1.5, and 1 mas --- using the `leaf' clustering method (Figure \ref{fig:comp}). The algorithm did not perform effectively for a cut-off parallax $>$10 mas, and thus, with exception of the Hyades, little of the structure within 100 pc can be recovered. 

\begin{figure*}
\epsscale{1.1}
\plotone{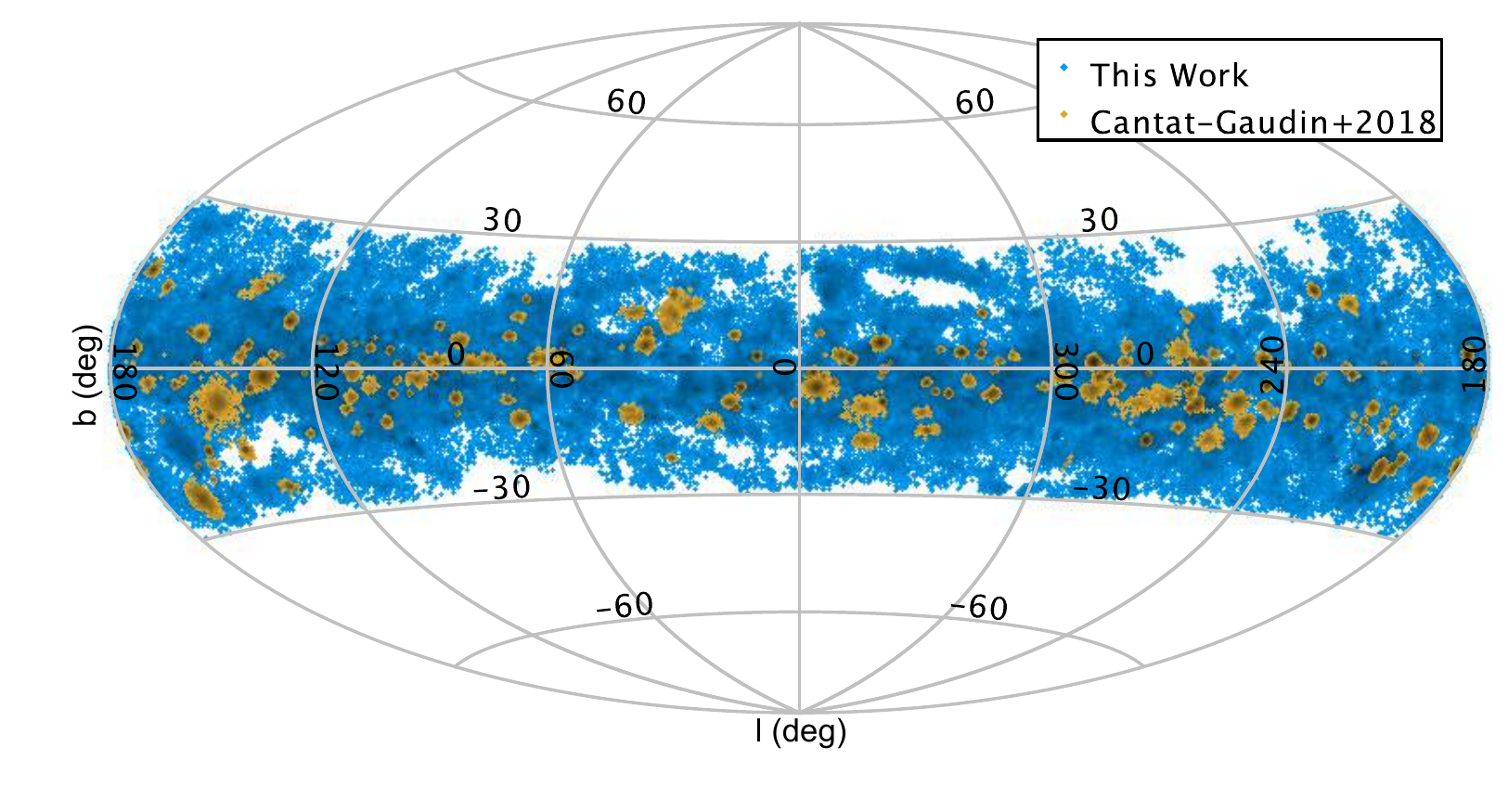}
\caption{Spatial distribution of sources in the final catalog in Galactic coordinates. Sources in common with \citet{cantat-gaudin2018a} are shown in yellow.
\label{fig:CG}}
\end{figure*}

Compared to the leaf method, the default `eom' (excess of mass) method better recovers extended lower density structure in various star-forming groups (e.g. Orion, Sco-Cen, Taurus, Perseus). In most cases, however, it would return only one or two group for the entire volume of space being probed, merging all of the substructure into a single coherent distribution of sources. This may occur due to a large number of outlying sources that do not belong to physically significant structures. Only a single optimal configuration of the input data was found at a cut-off parallax of 2.0 mas that did not result in this overmerging. It was added as a supplement to the runs performed with the leaf method, which had a significantly more consistent performance in preserving distinct structures.

The Galactic coordinate grid is discontinuous at $l=0=360^\circ$. Thus, some of the structures that cross that boundary are artificially split (e.g. Figure \ref{fig:comp}). Thus multiple runs are performed for each parallax cut, spanning from $0^\circ<l<360^\circ$, and from $-180^\circ<l<180^\circ$ up to 2 mas. Because of the number of sources necessary to process, for computational efficiency, runs at 1.5 and 1 mas are split to run from $0^\circ<l<190^\circ$, and from $-180^\circ<l<10^\circ$.

Merging outputs of different runs with the same cut-off parallax (i.e., to stitch up the discontinuities in $l$) is largely a trivial process. In some of the overlapping clusters, a handful of sources might be absent in one run compared to another due to slight inconsistencies in the weights given by HDBSCAN with different input matrices. But they are tracing the same underlying structures, and most of them are identical, with primary differences occurring at the location of the split in $l$.

On the other hand, merging outputs from different parallax runs is somewhat more complex. With different density sensitivity, it is possible that in a particular run, some sources in neighboring structures might be clustered into one group, whereas these structures might be split in into distinct structures in a run sampling a different volume. All of the outputs from different cut-off parallax runs were examined for self-consistency of their position on the sky, proper motions, parallaxes, and the shape of the HR diagram. They were merged manually, and split by hand if there were natural divides in any of the aforementioned spaces. Some of the obvious contamination in individual groups was also removed -- this was most noticeable in young clusters located far above the main sequence on the HR diagram that had a small fraction of more evolved stars that appeared to be uniformly distributed in all of the position-velocity parameter spaces. This amounted to $\sim$3000 stars, $\sim$1\% of the total sample.

It should be reiterated, however that there is a great degree of complexity in the physical and kinematical structure even in the young star forming regions. While they can be roughly broken apart into smaller components of varying densities that may evolve separately over time, they are fundamentally formed from the same parental cloud complex, and many of these divisions may be somewhat arbitrary. In Orion, for example, the Complex can be subdivided into individual clouds; each cloud can then be subdivided into individual clusters and diffuse regions; the clusters themselves may be subdivided into subclusters and other notable features \citep{kounkel2018a}. There is no unique method with which all of the different scales of structure could be split into individual components, and the boundaries between them are fuzzy \citep[e.g.,][]{beccari2017,zari2017,grosschedl2018,kuhn2019,getman2019,chen2019}. This may also the case for many of the individual groups identified in this work, as will be discussed in later sections.

The final catalog totals 1,901 individual groups consisting of 288,370 stars (Table \ref{tab:data}). Combined, these 288K sources that were grouped together represent less than 1.5\% of the $\sim$20 million stars in the original \textit{Gaia} DR2 input catalog. The average properties of the resulting groups (further processed as described in Section \ref{sec:string}; hereafter named ``Theia''), are given in Table \ref{tab:cluster}. 

Comparing this catalog to the catalog of open clusters identified in \textit{Gaia} DR2 by \citet{cantat-gaudin2018a}, we recover 198 clusters out of 208 that are located within 1 kpc; within those we recover 73\% of the sources identified as members of those clusters. However, the catalog presented in this work is significantly more sensitive to the extended structure, recovering not only gravitationally bound clusters, but also a number of comoving groups and associations, most of which have not been previously known (Figure \ref{fig:CG}).

We estimate that as much as 5--10\% of the population inside of the identified groups may be contamination, consisting of field stars that have kinematics similar to those of the underlying cluster. Additionally, while it is difficult to estimate the precise fraction, there may also be some number of groups consisting of unrelated stars that appear to be co-moving. Their fraction is expected to increase at larger distances. RVs from \textit{Gaia} can confirm some of the structures, particularly those that are extended, such as those described in Section \ref{sec:string}. However, as \textit{Gaia} does not have RVs for all the stars in the Theia catalog, and the uncertainties for the existing measurements are large, it is difficult to determine which one of the more compact groups are real and which one are spurious. Future spectroscopic follow-up would be needed to distinguish between them based on coherence in RV and chemical composition. Nonetheless, even ``fake'' clusters can be used as tracers of the underlying structure and kinematics of the extended solar neighborhood.

\section{Ages} \label{sec:ages}

In order to estimate the ages of the individual groups, two approaches were considered.

\subsection{Machine Learning}

To infer ages for each group identified in this analysis, we constructed a convolutional neural network (CNN) using PyTorch \citep{pytorch}. The CNN was trained on the cluster populations from \citet{cantat-gaudin2018a}, with the age and extinction labels from \citet{kharchenko2013}, with the combination of the \textit{Gaia}, 2MASS, and WISE photometry which were obtained from the cross match tables \citep{marrese2019}. Each individual cluster in the training set was subsampled repeatedly with a random number of sources from 40 to 250 stars, to increase the sample from 1196 clusters to 130,000 realizations of them. Even though the same stars were represented in many different realizations, because their combination was slightly different, it allowed the CNN to more effectively memorize the different features of clusters with various ages than it could have from from a smaller set of clusters with more complete sampling of their stellar populations.

Additionally, the training set was supplemented by population drawn from synthetically generated clusters. We generated clusters with a uniform distribution of ages (in log space) from 1 Myr to 10 Gyr with a 2 Myr scatter within a cluster. The clusters' remaining properties were randomly generated as well, with uniform Fe/H from -2.5 to 0.5, uniform distances from 50 to 3000 pc with a 2 pc scatter, uniform $A_V$ from 0 to 2 with 0.05 scatter, and the masses of the individual stars were drawn from the initial mass function. \textit{Gaia} DR2 ($G$, $BP$, $RP$), 2MASS ($J$, $H$, $K$) and WISE ($W1$, $W2$, $W3$) fluxes were interpolated using the PARSEC isochrones \citep{marigo2017}, and the typical uncertainties in each band (as well as in the parallax) were applied. We included only the fluxes brighter than $G<19$, $BP<20.5$, $RP<17.5$, $J<17$, $H<16$, $K<16$, $W1<16$, $W2<16$, $W3<13.5$, which are the typical limiting magnitudes in our observed sample. Only the $G$ band is required for all the synthetic sources; if other bands are undetected they were set to the edge value. A total of 150,000 synthetic clusters were generated to supplement the real observational data, in which the missing fluxes were processed in the same manner.

All the sources in the individual clusters were ordered based on their $M_G$, and all of the 9 fluxes and the parallax were passed to the CNN which consisted of 7 2d-convolutional and 2 fully connected layers, architecture of which is shown in Appendix \ref{sec:cnn}. 

\subsection{Isochrone fitting}

In addition to the machine learning approach, we estimate the ages of the individual groups through traditional isochrone fitting. We used the Bayesian Analysis for Stellar Evolution with nine variables \citep[BASE-9,][]{von-hippel2006,base9}. Inputs for BASE-9 fitting were the absolute photometry from the $Gaia$ filters, incorporating $\sigma_\pi$ as part of the photometric uncertainties, with a minimum photometric uncertainty of 0.02 mag. PARSEC isochrones \citep{marigo2017} were used for fitting, although their current implementation in BASE-9 cannot estimate ages younger than 7.4 dex. For computational expediency, binaries were not treated in the fitting process, and various combinations of the input parameters and chain lengths were considered. However, due to the volume of data, the resulting chains were not evaluated on whether the output was correlated or not.

\subsection{Comparison and the final parameters}

\begin{figure}
\epsscale{1.1}
\plotone{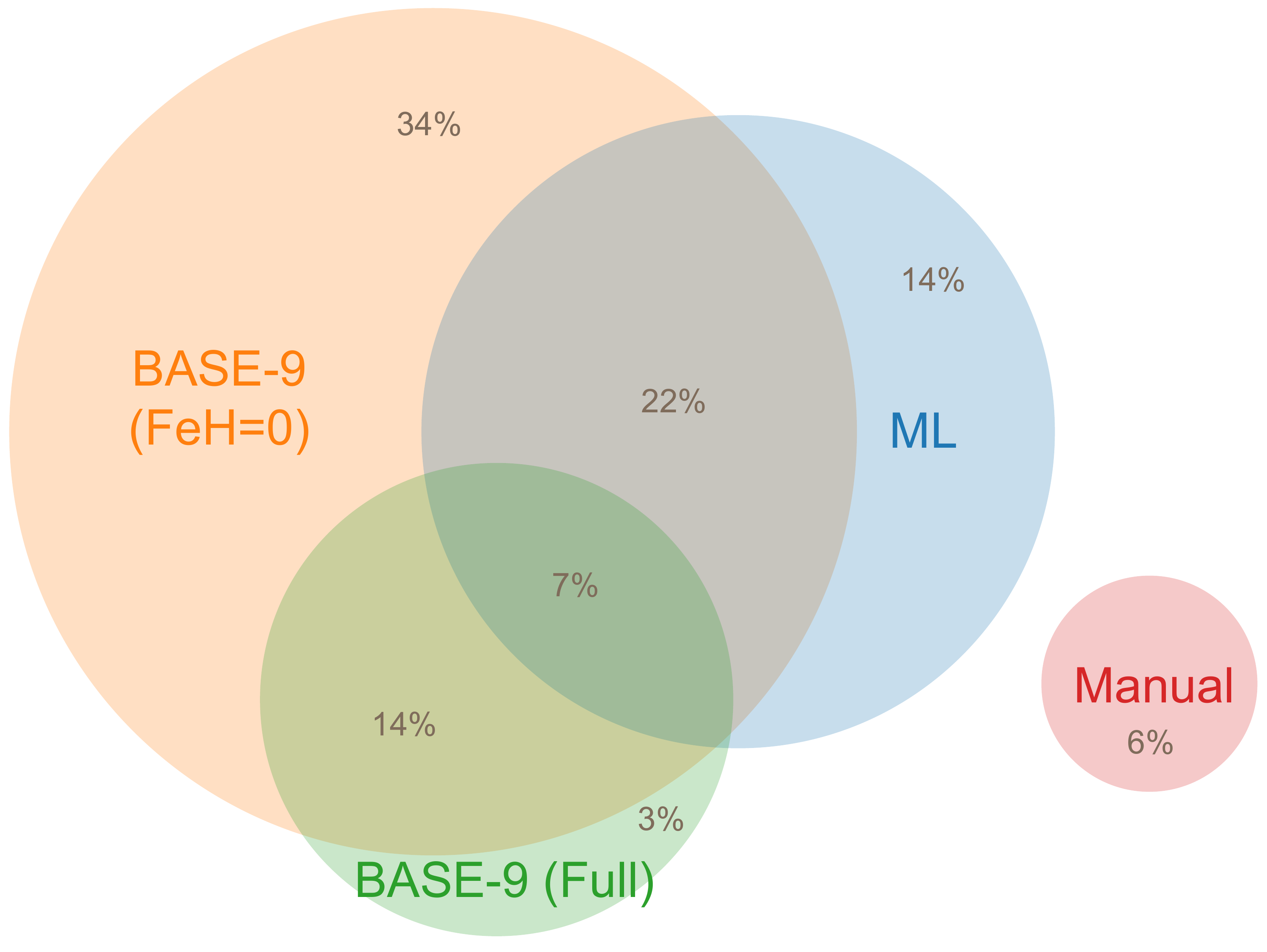}
\caption{Distribution of success cases of various methods to determine age and extinction.
\label{fig:chart}}
\end{figure}

\begin{figure}
\epsscale{1.1}
\plotone{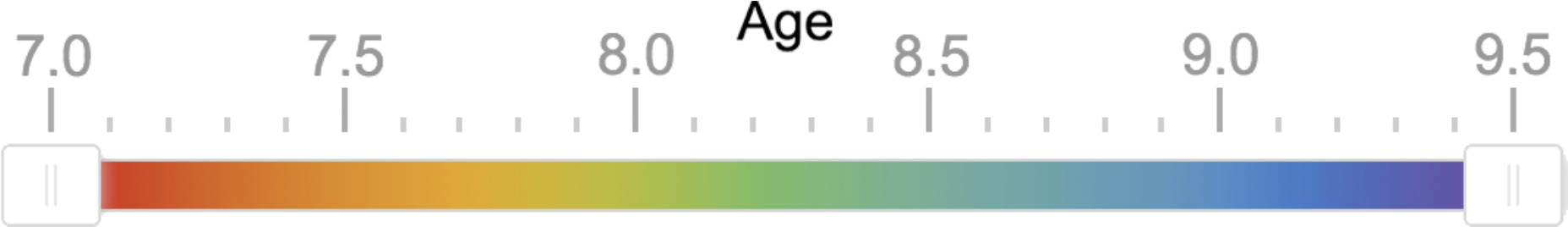}
\plotone{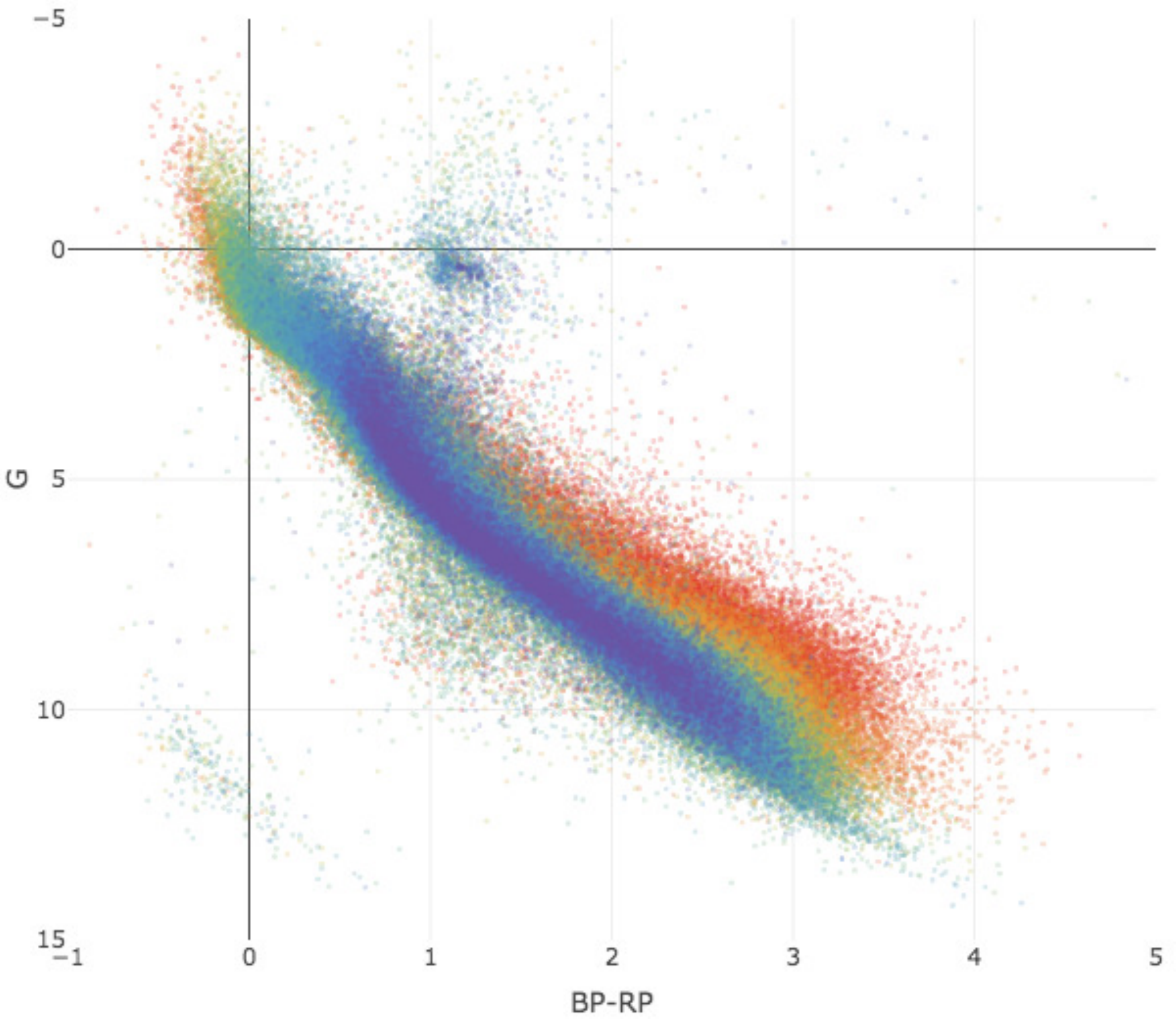}
\caption{Extinction-corrected HR diagram of all the sources, color-coded by their measured age. An interactive version of the figure, with a combination of various photometric bands, is available in the online version. (Temporarily at \url{http://mkounkel.com/mw3d/hr.html})
\label{fig:hr}}
\end{figure}

\begin{figure}
\epsscale{1.1}
\plotone{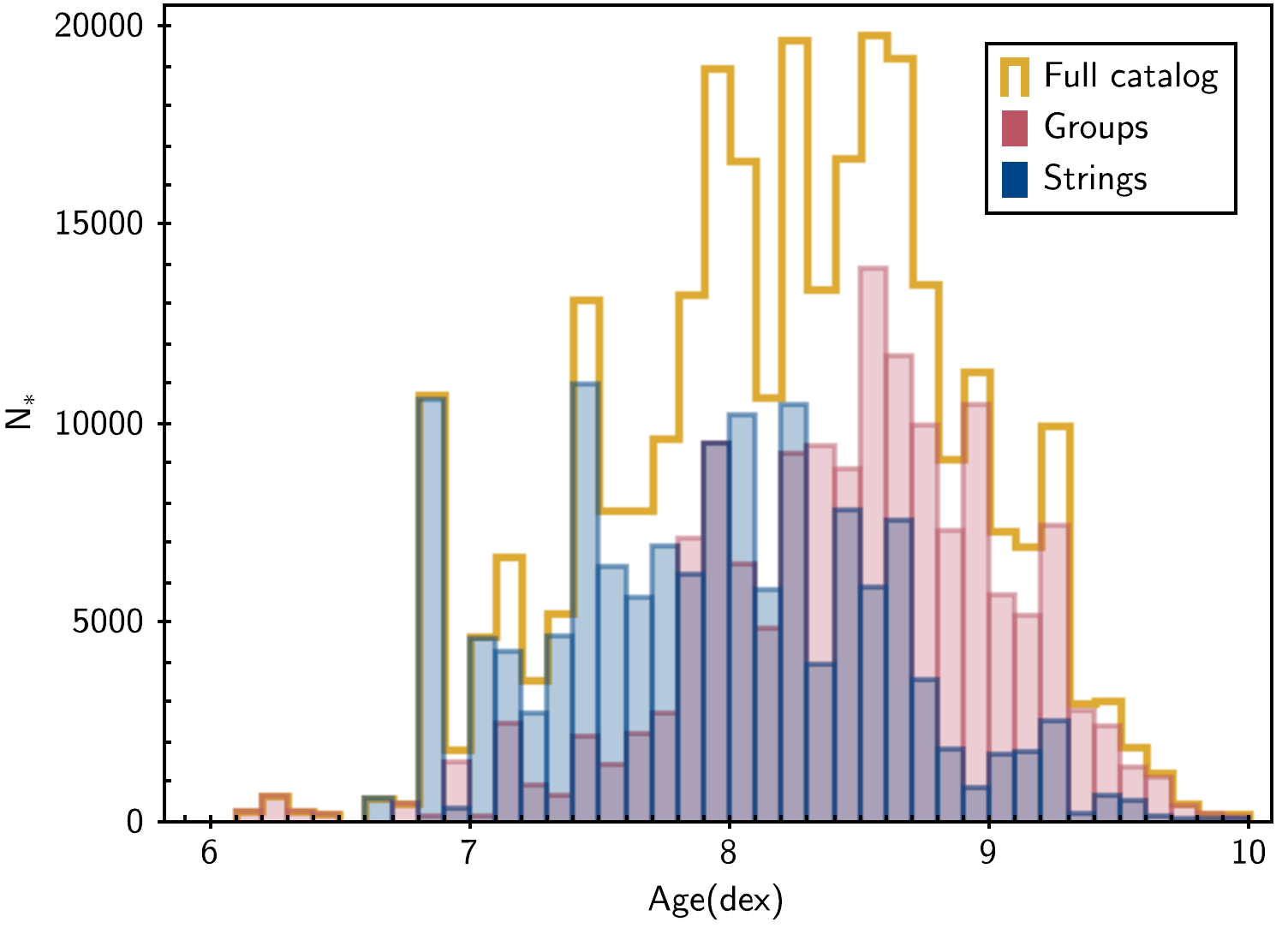}
\caption{Distribution of stellar ages in the catalog of the coherent structures.
\label{fig:age}}
\end{figure}

\begin{figure}
\epsscale{1.1}
\plotone{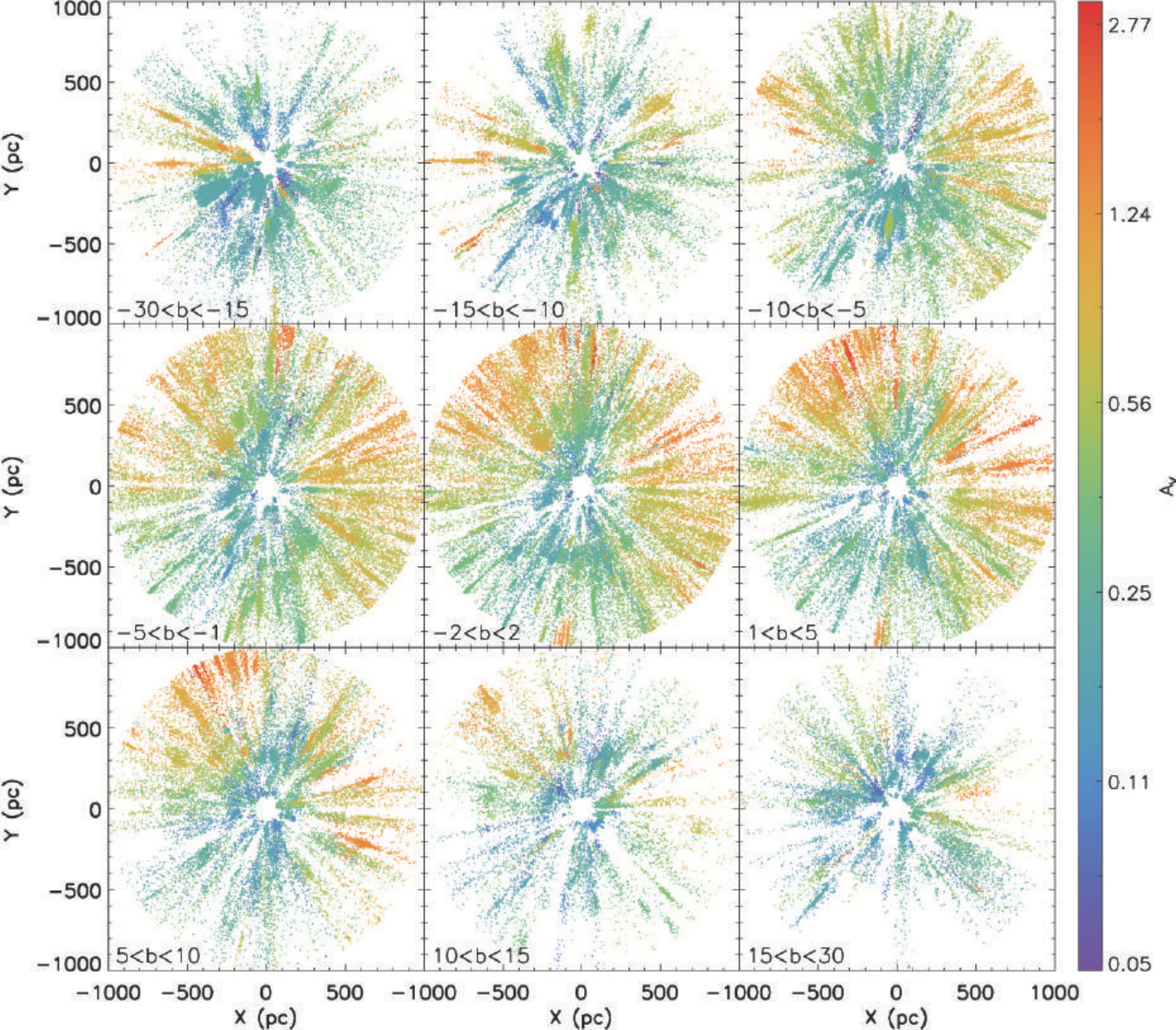}
\caption{Map of the measured extinction, Galactic center is towards the right of the plot
\label{fig:av}}
\end{figure}

The resulting isochrones, constructed both from the parameters predicted by CNN and those fitted using BASE-9, were visually examined against a color-magnitude diagram for each of the identified groups. Multiple outputs were averaged together if they produced a reasonable approximation of the data. The most successful approach was the isochrone fitting restricted to solar metallicity, not fitting for a potential offset in the distance modulus due to systematic of the \textit{Gaia} parallaxes \citep[e.g.,][]{stassun2018}, and using the predictions from CNN as the initial estimates for age and $A_V$. This produced a successful fit in 77\% of cases. Machine learning on its own produced reasonable approximation in 44\% of cases. BASE-9 fitting with additional parameters such as distance and metallicity could approximate isochrones for only 24\% of groups (without additional tuning of the inputs). And 6\% of groups struggled to be characterized by either method and required manual fitting (Figure \ref{fig:chart}).

From the scatter in the accepted parameters we estimate the typical uncertainties in age to be $\sim$0.15 dex, and $\sim$0.2 mag in $A_V$. This is a sufficient degree of precision for the purposes of the analysis in this work, however, these age estimates do not supersede those from the more detailed studies of individual clusters, and more work will be needed in the future to improve on these estimates, such as through gyrochronology. The primary origin of the uncertainty in our age determination originates from a lack of a distinct turn off of the top of the main sequence to the red giant branch, as well as over-inflated low mass stars compared to the isochrones in some of the populations \citep[e.g.][]{jackson2018}.  These uncertainties partially reflect stochastic effects, due to both intrinsic astrophysical fluctuations in the cluster's population and subsampling by the fitting routines, but also competing systematic biases (i.e., where a missing turnoff star will drive the cluster to older ages, whereas the mass dependent offsets from isochrones at lower masses drive to younger age estimates). However, we do note that the ages were derived independently of any analysis of the galactic positions or kinematics (Section \ref{sec:string}). The apparent continuous evolution of the populations in time and their consistency with the Galactic kinematics do demonstrate that the derived ages are reliable and self-consistent. Furthermore, the overall position of the sources on the HR diagram (Figure \ref{fig:hr}), does also show a smooth gradient as a function of age.

While using machine learning to predict cluster parameters is still inferior to the direct fitting method, it does show  potential, particularly as it requires fewer computational resources to process large quantities of data. A fully trained network can evaluate the parameters in the entire dataset in a matter of seconds. At the same time, while BASE-9 utilizes a Bayesian approach in evaluating the parameters through a Markov chain Monte Carlo, the quality of the fit is dependent on the input parameters. And, in particular, BASE-9 appears to run just a single long chain with a very limited number of walkers, which can be sub-optimal compared to multiple walkers in probing the full parameter space \citep{goodman2010,foreman-mackey2013,bian2017}. While it is possible to run BASE-9 multiple times to produce multiple chains, as mentioned before, due to the volume of data it was impractical to do.

The overall distribution of the measured ages is shown in Figure \ref{fig:age}, and further discussed in the subsequent sections. The 3-d distribution of the measured $A_V$ is shown in Figure \ref{fig:av}; it is largely self-consistent (i.e., increasing with distance along the line of sight), and the overall distribution is consistent on a per-cluster basis with what is mapped by the extinction measurements of individual sources from \textit{Gaia} DR2.

\section{Strings}\label{sec:string}
\subsection{Manual assembly}
\begin{figure*}
\epsscale{1.1}
 \centering
		\gridline{\fig{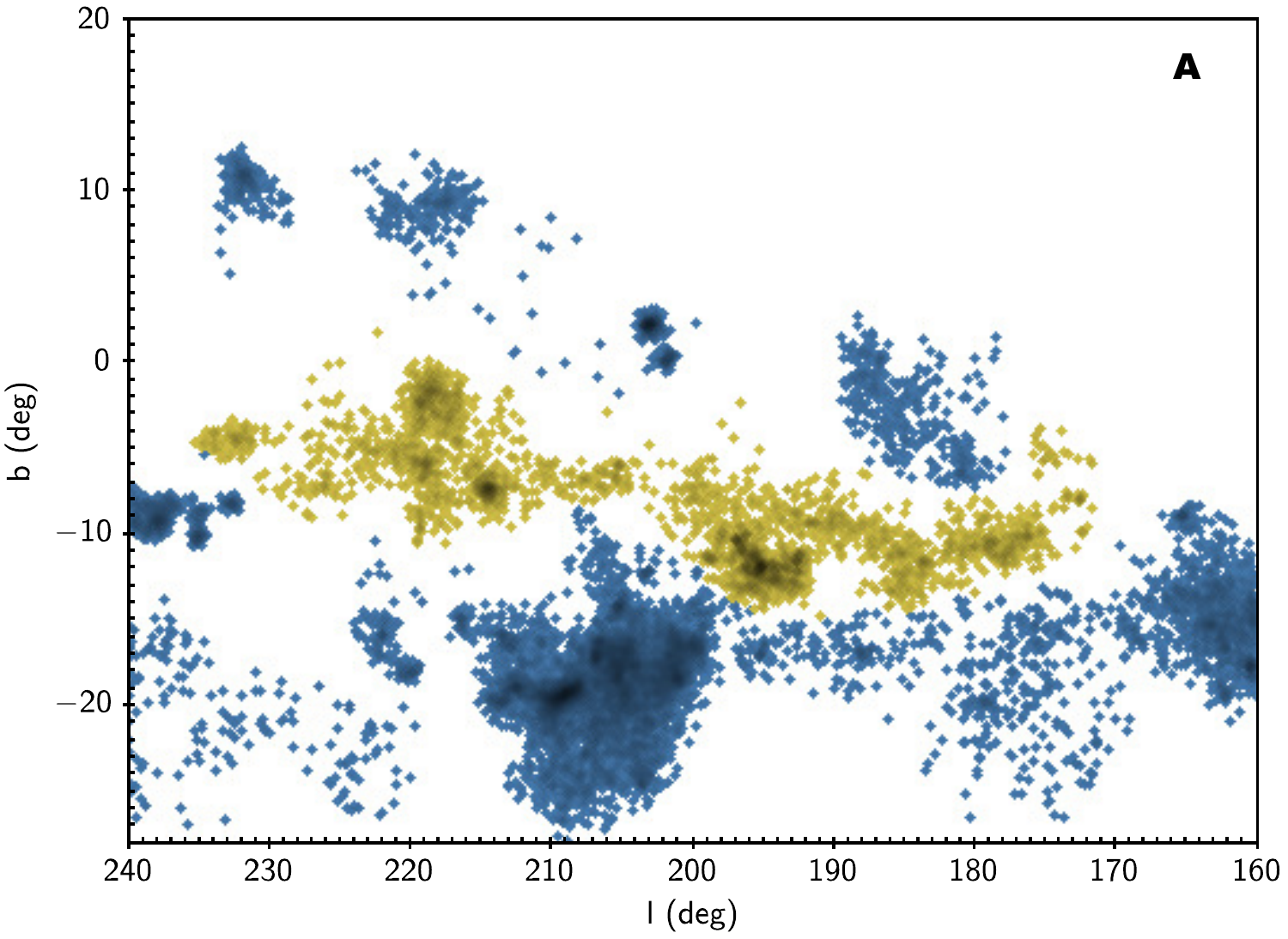}{0.33\textwidth}{}
             \fig{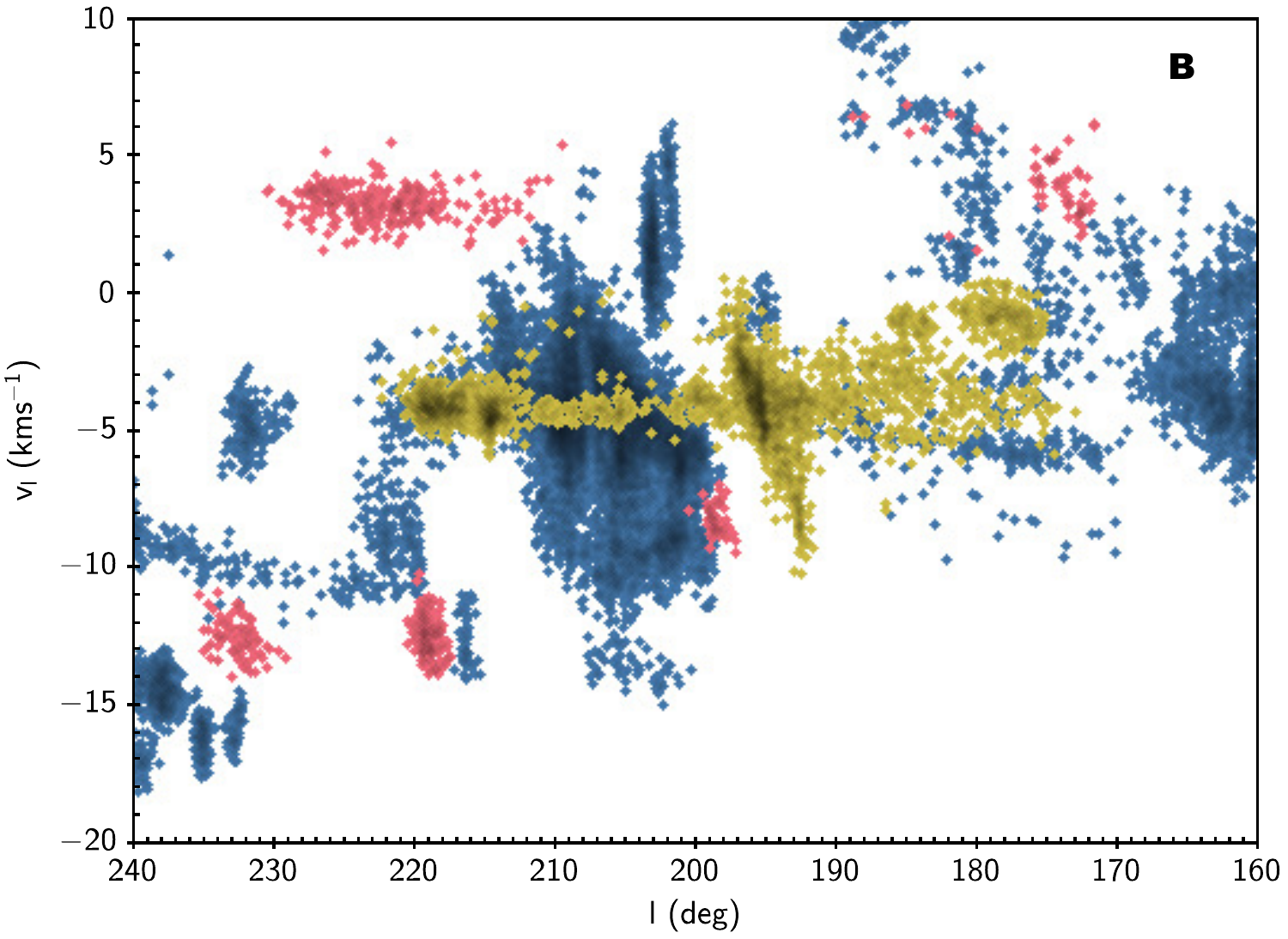}{0.33\textwidth}{}
             \fig{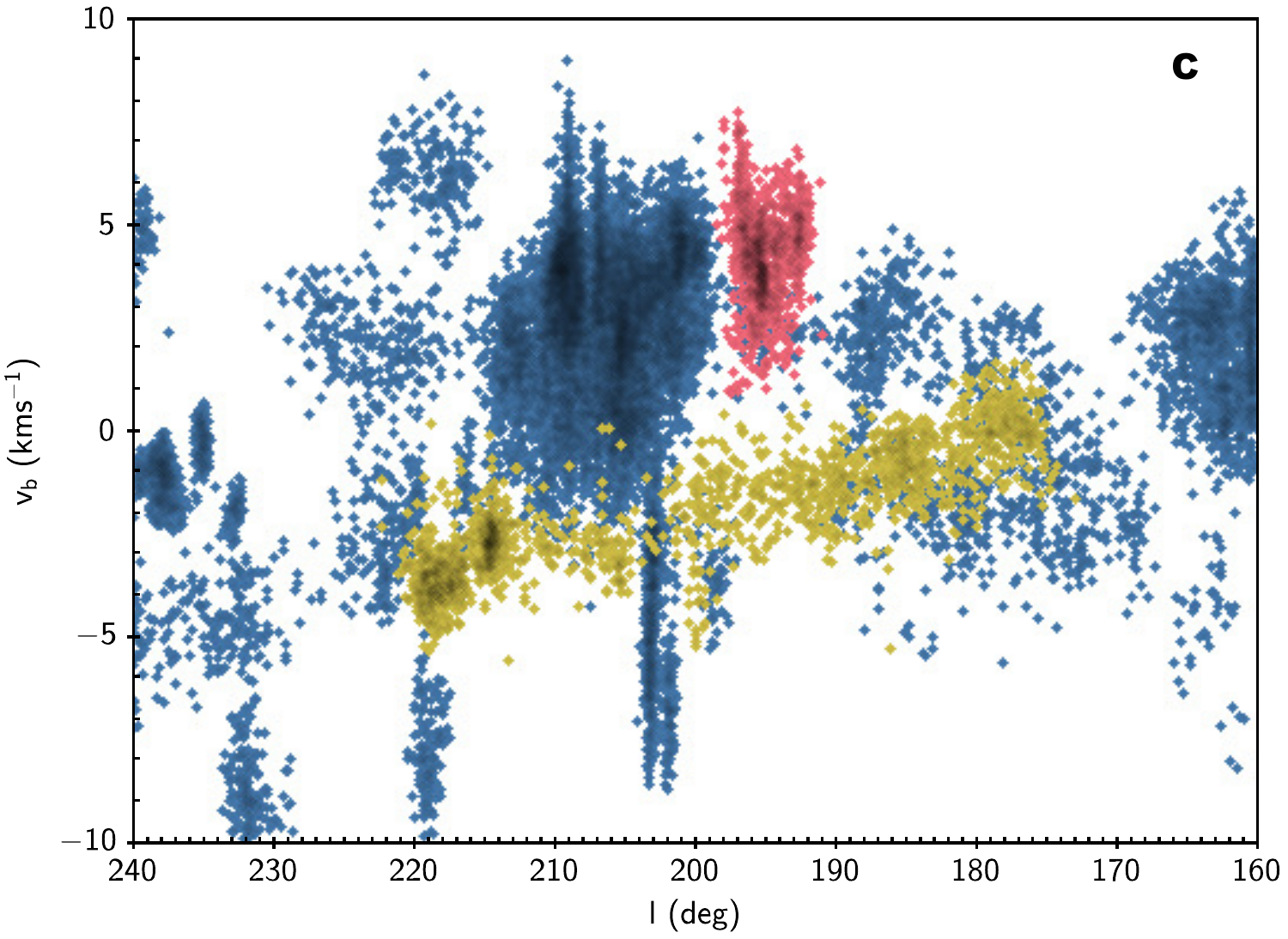}{0.33\textwidth}{}
        }\vspace{-0.75 cm}
		\gridline{\fig{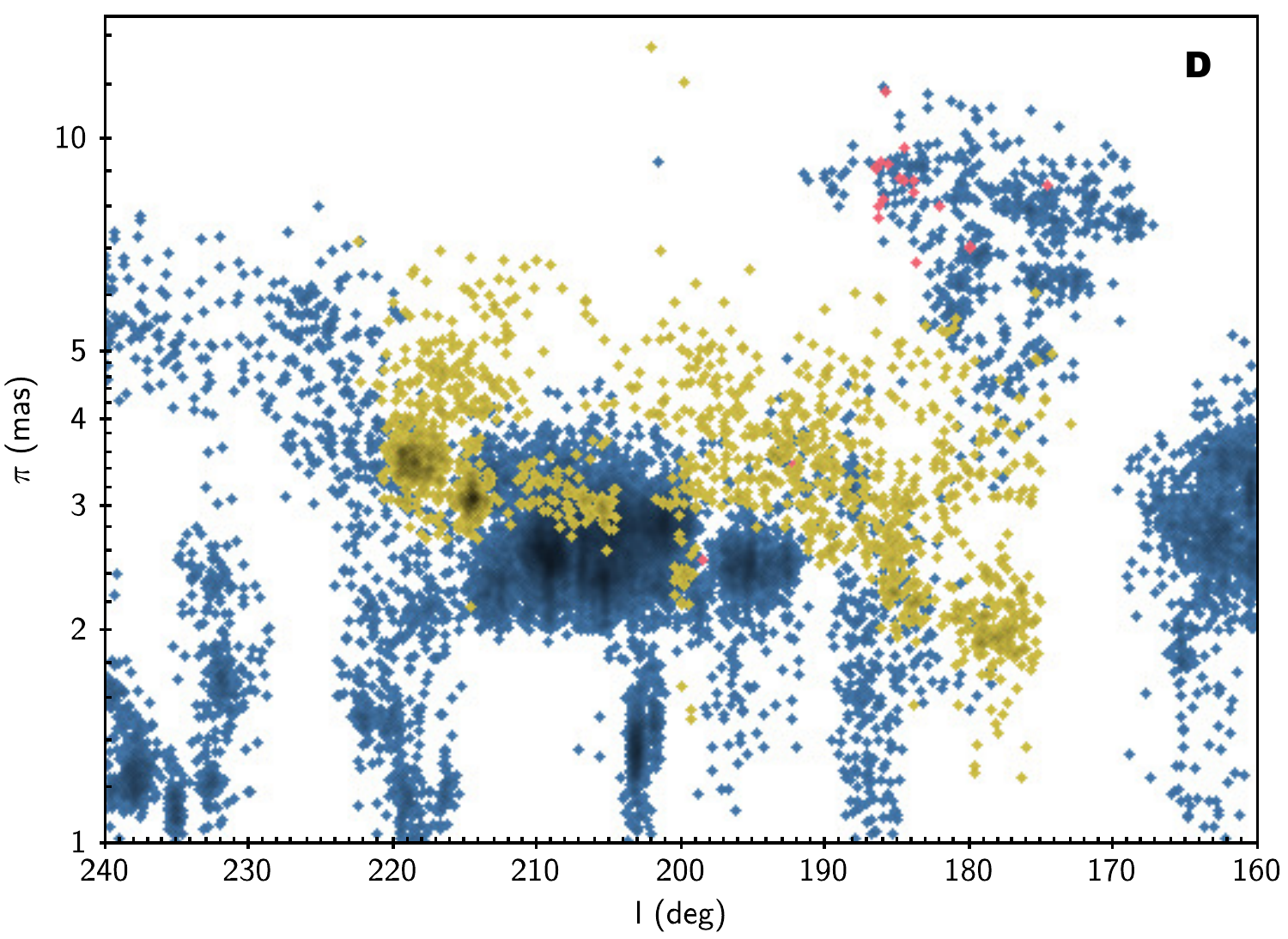}{0.33\textwidth}{}
             \fig{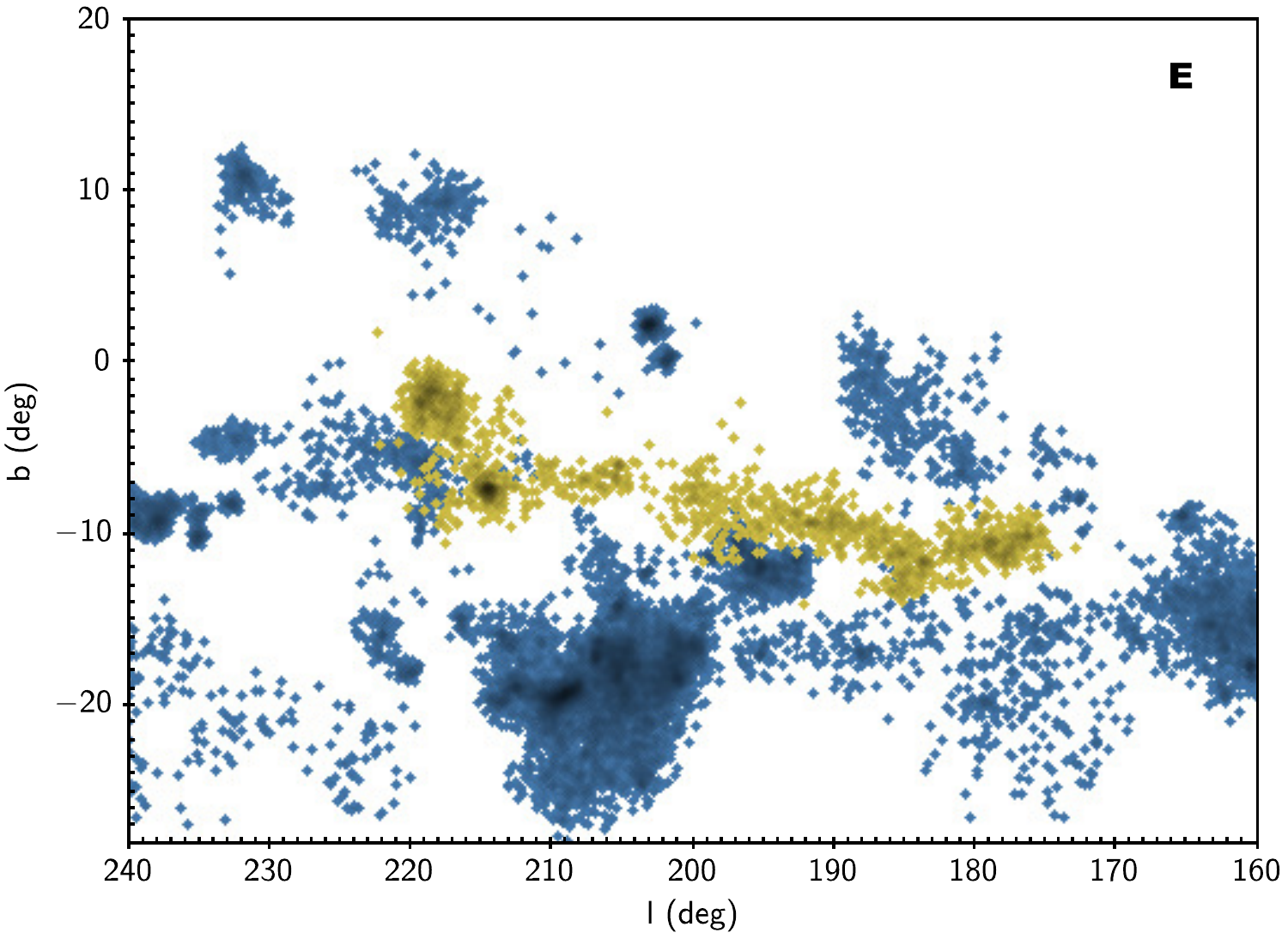}{0.33\textwidth}{}
             \fig{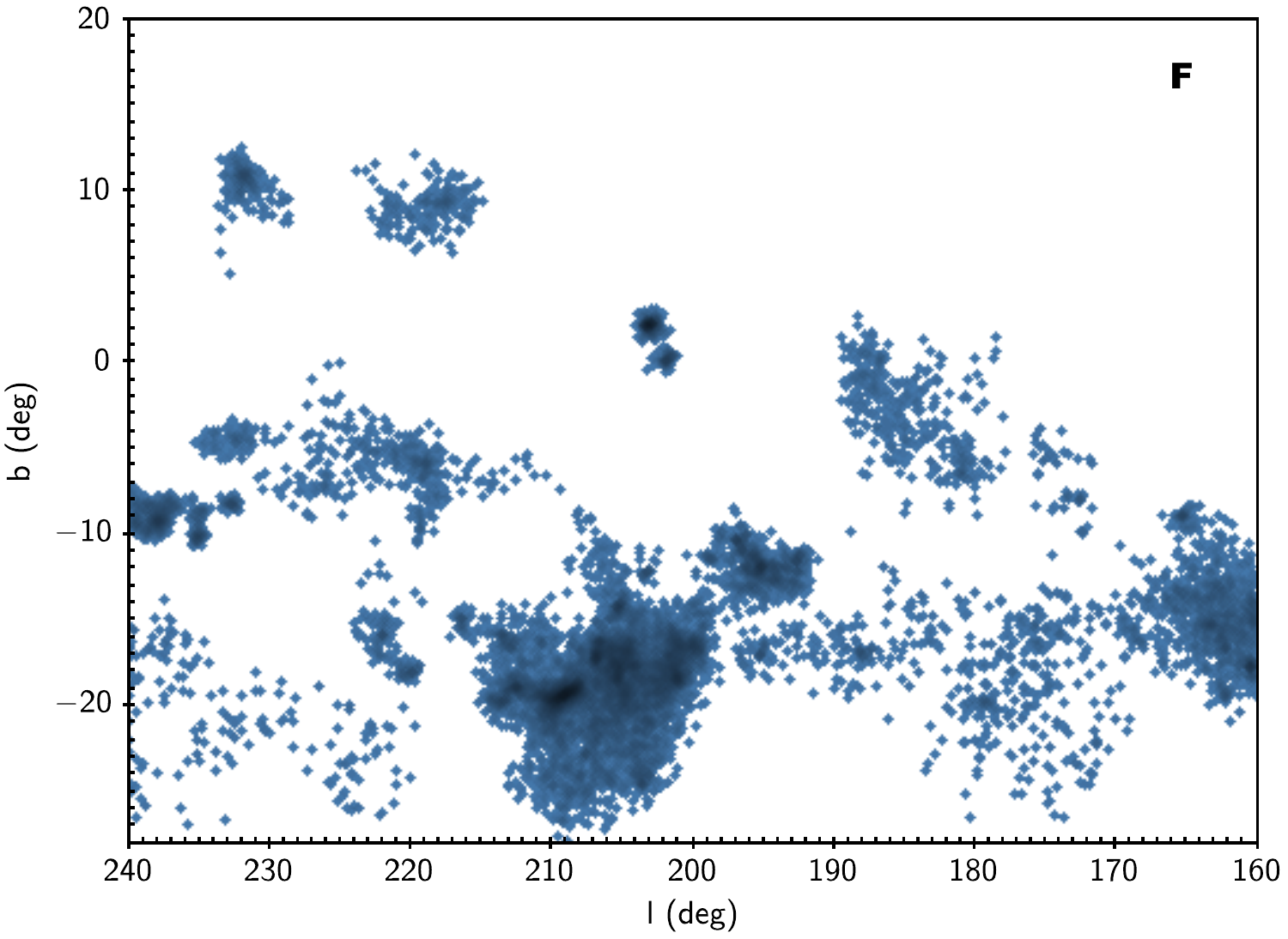}{0.33\textwidth}{}
        }\vspace{-0.75 cm}
\caption{A demonstration of the selection of a string. NGC 2232 is shown as an example. Blue colors correspond to the full catalog for ages of up to 7.5 dex. Sources selected at the current step are shown in yellow, while the sources selected at the previous step and discarded at the current step are shown in red. A: rough selection of a string in $l$ vs $b$. B: refinement in $l$ vs $v_l$, removing sources with distinct, non-continuous $v_l$ kinematics. C: refinement in $l$ vs $v_b$. D: refinement in $l$ vs $\pi$. E: the resulting selection in $l$ vs $b$ space, for comparison with the panel A. F: the remaining catalog, with the groups that compose the string removed.
\label{fig:select}}
\end{figure*}

The distribution of many of the identified groups appears to form filamentary structures that we will refer to as strings. These strings are roughly parallel to the Galactic plane, they appear to be coherent both spatially and kinematically, and the stars in them generally can be characterized with a single isochrone (with a possible scatter of few Myr). In some cases, the entire identified group may correspond to a single such string. In other cases, multiple groups with similar ages appear to form spatially extended but kinematically coherent strings, linear extension of which was segmented by HDBSCAN.

To identify and manually reconstruct these strings that consist of multiple groups, we have examined the distribution of sources in $l$ vs. $b$, $l$ vs $\pi$, $l$ vs $v_l$, and $l$ vs $v_b$ and over a range of age slices (e.g., Figure \ref{fig:select}). Using TOPCAT \citep{topcat}, we selected a structure that appears to be most well-defined and separatable from its surroundings in one of these planes. We compared their coherence in the other planes, removing outlying sources in each one, until the remaining structure was fully continuous coherent in all kinematic and spatial dimensions. We verified that the manual selection was representative of the groups to which these sources belong, in that individual groups were not split into two, and most of their sources were recovered with the selection. These groups were then joined into a common string. If, instead of multiple groups, only a single filament-like group remained after the selection process, its classification was also assigned to a string. Such strings that consisted of a single group typically have comparable filamentary shape to what is found in multiple group strings. The search was continued after removing the sources inside the identified string from the catalog that would be used in the next iteration of the search process. This process was able to untangle multiple structures that were projected on top of each other in any of the dimensions. It should be noted that a union of some individual clusters may appear to form coherent structures in some projections of the phase space, but they tend to become incoherent in the remaining projections. No grouping was applied in such cases.

Afterwards, we examined the remaining individual groups that have not been joined into strings, searching for extended structure with a strong eccentric distribution between $l$ and other dimensions, and they were similarly classified as strings.

Radial velocities were not used in identification of the strings due to their smaller sample size and poorer quality, although it should be noted that they tend to show a similar arrangement as a function of $l$ as can be observed in other dimensions. Once a more complete RV survey is available, however, the coherence of all the identified structures should be further revisited with the knowledge of the full 3d space motions.

\begin{figure*}
\epsscale{1.1}
 \centering
		\gridline{\fig{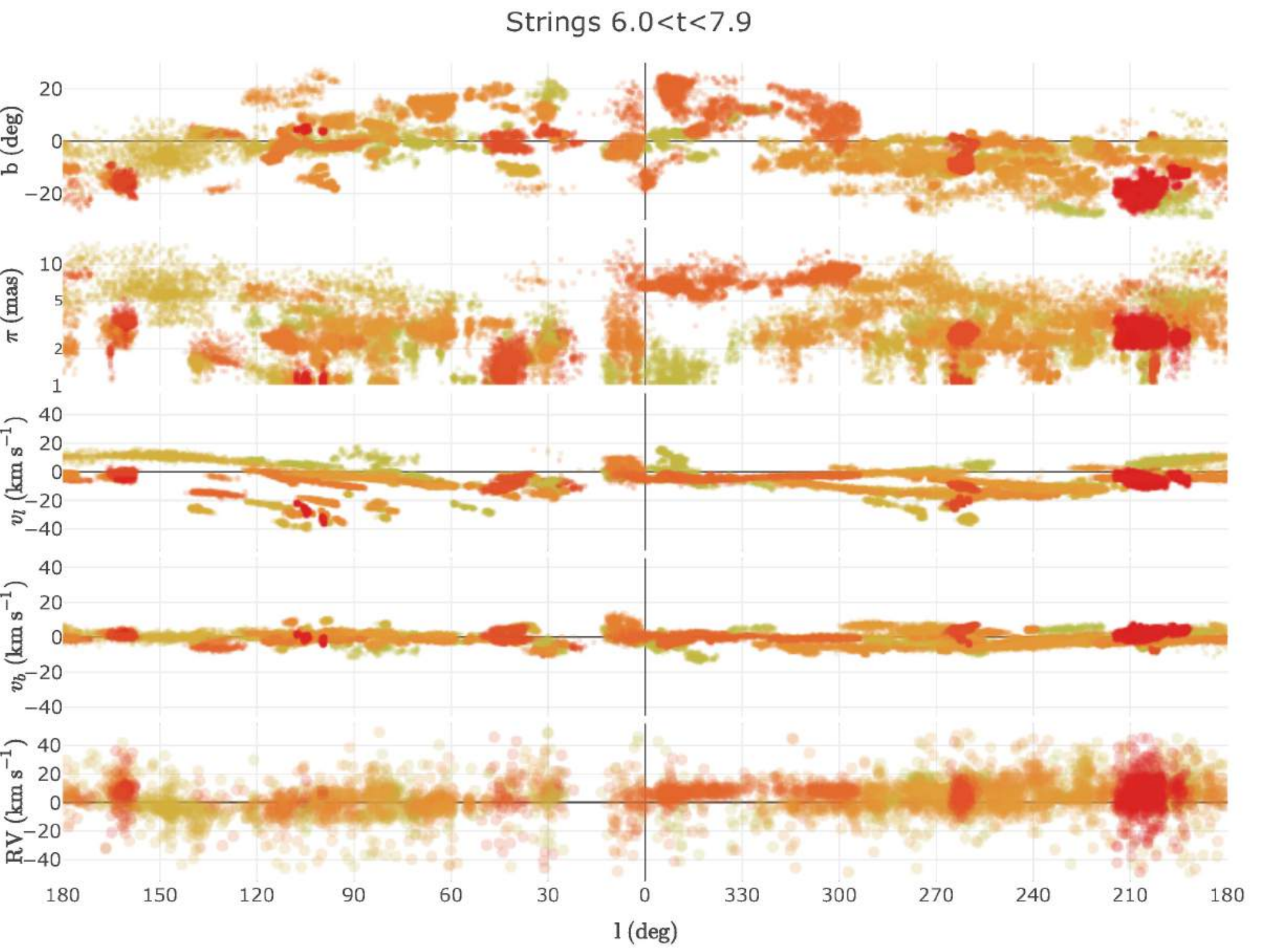}{0.5\textwidth}{}
             \fig{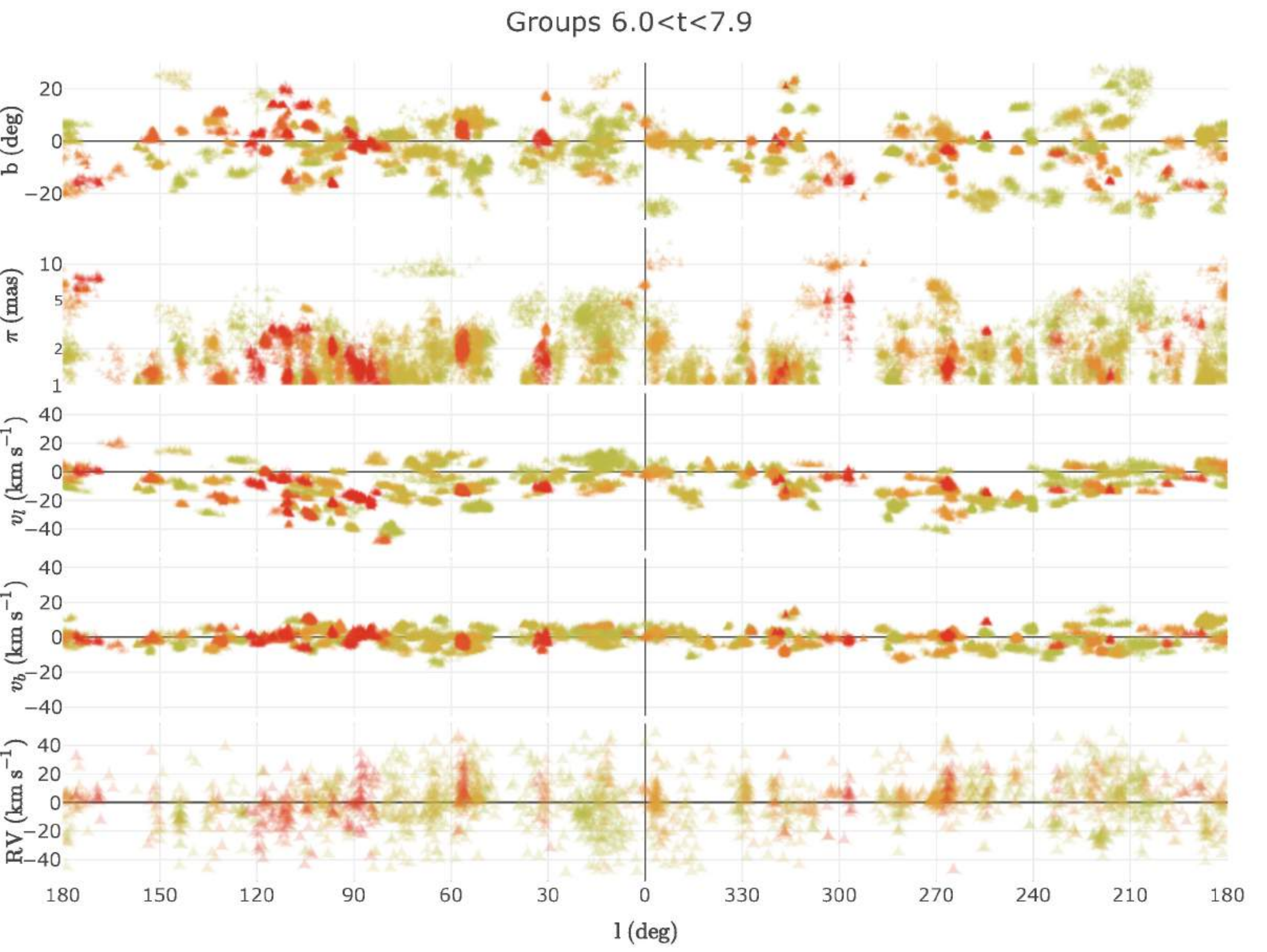}{0.5\textwidth}{}
        }\vspace{-1 cm}
		\gridline{\fig{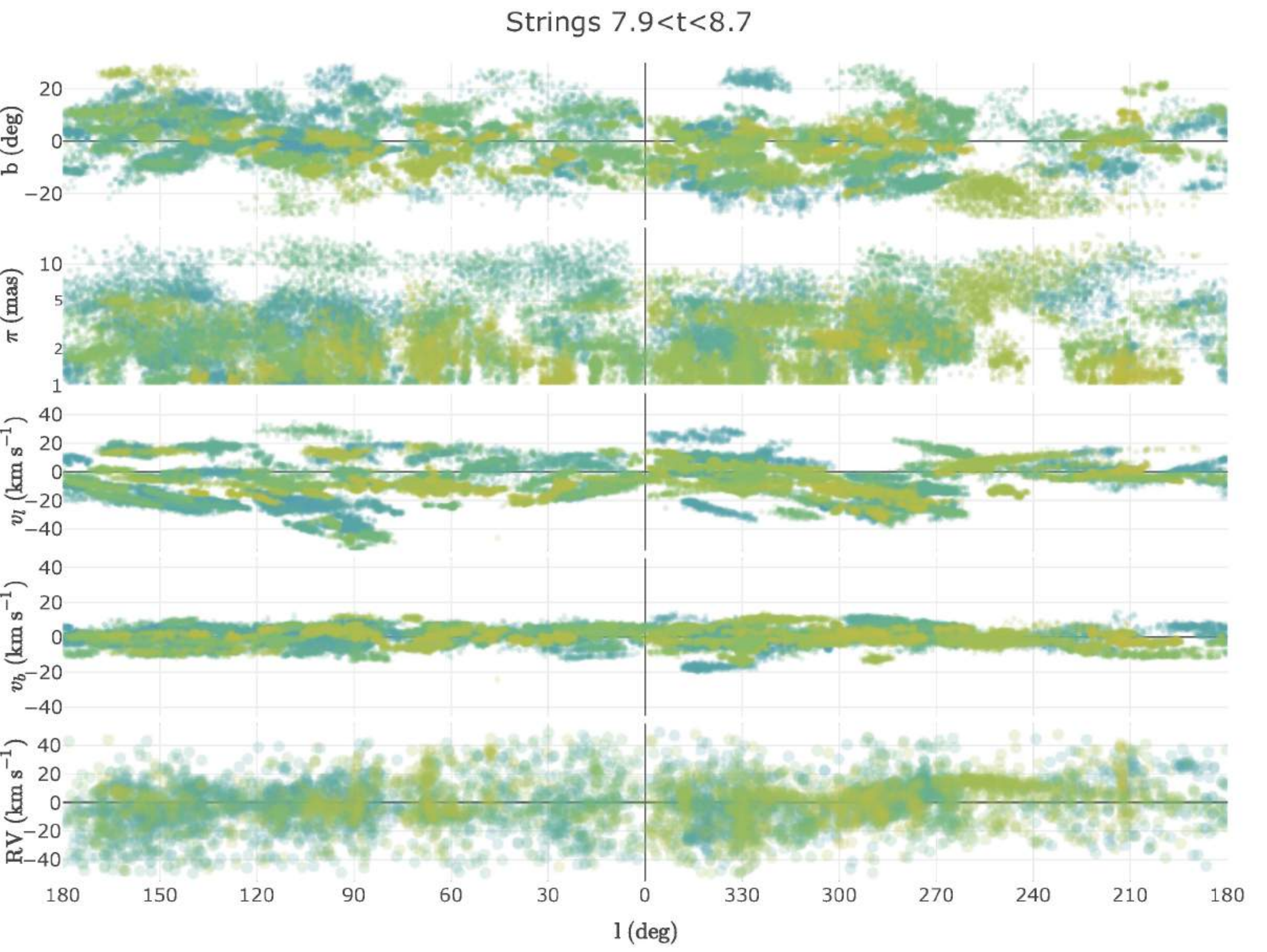}{0.5\textwidth}{}
             \fig{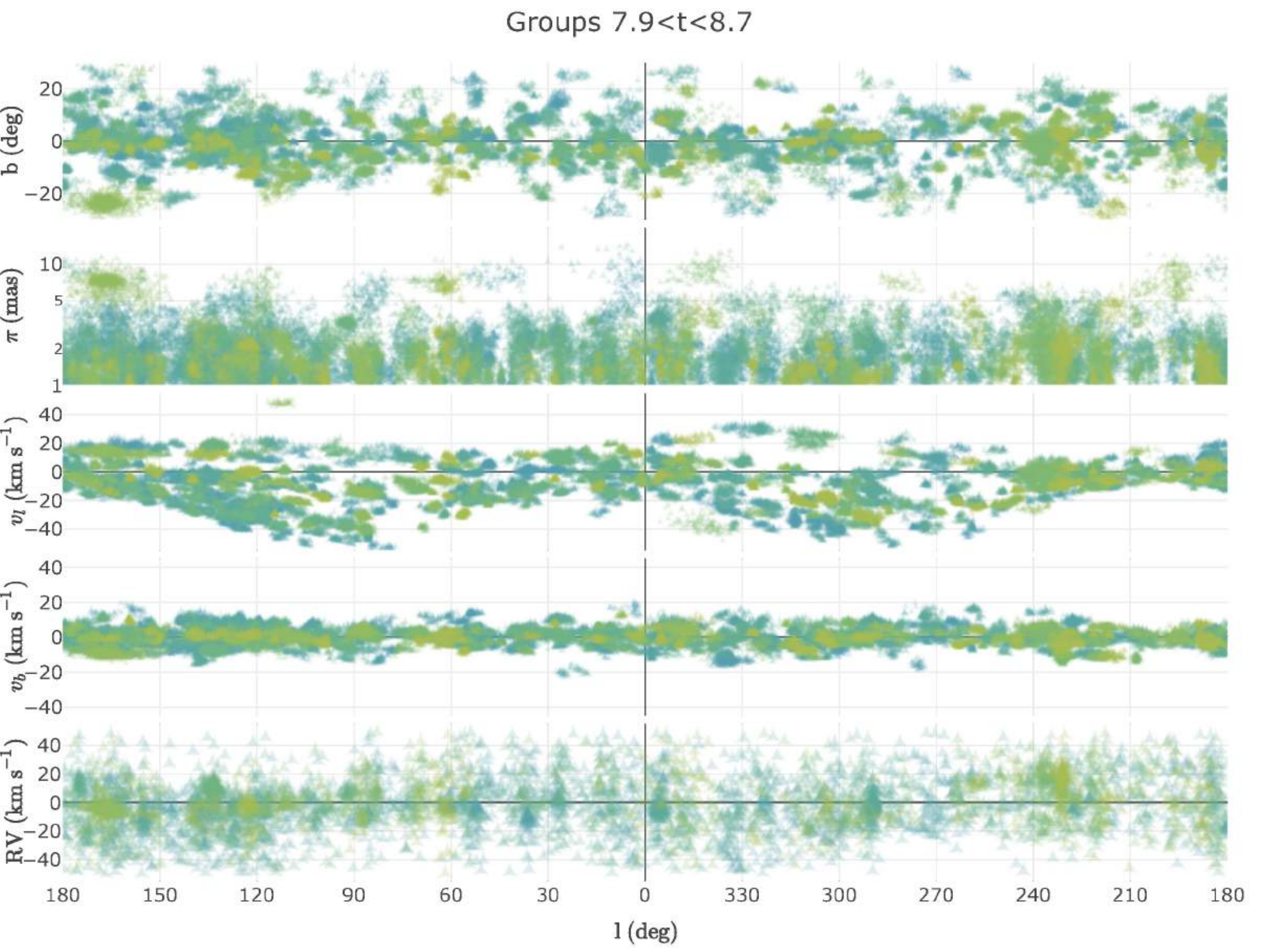}{0.5\textwidth}{}
        }\vspace{-1 cm}
		\gridline{\fig{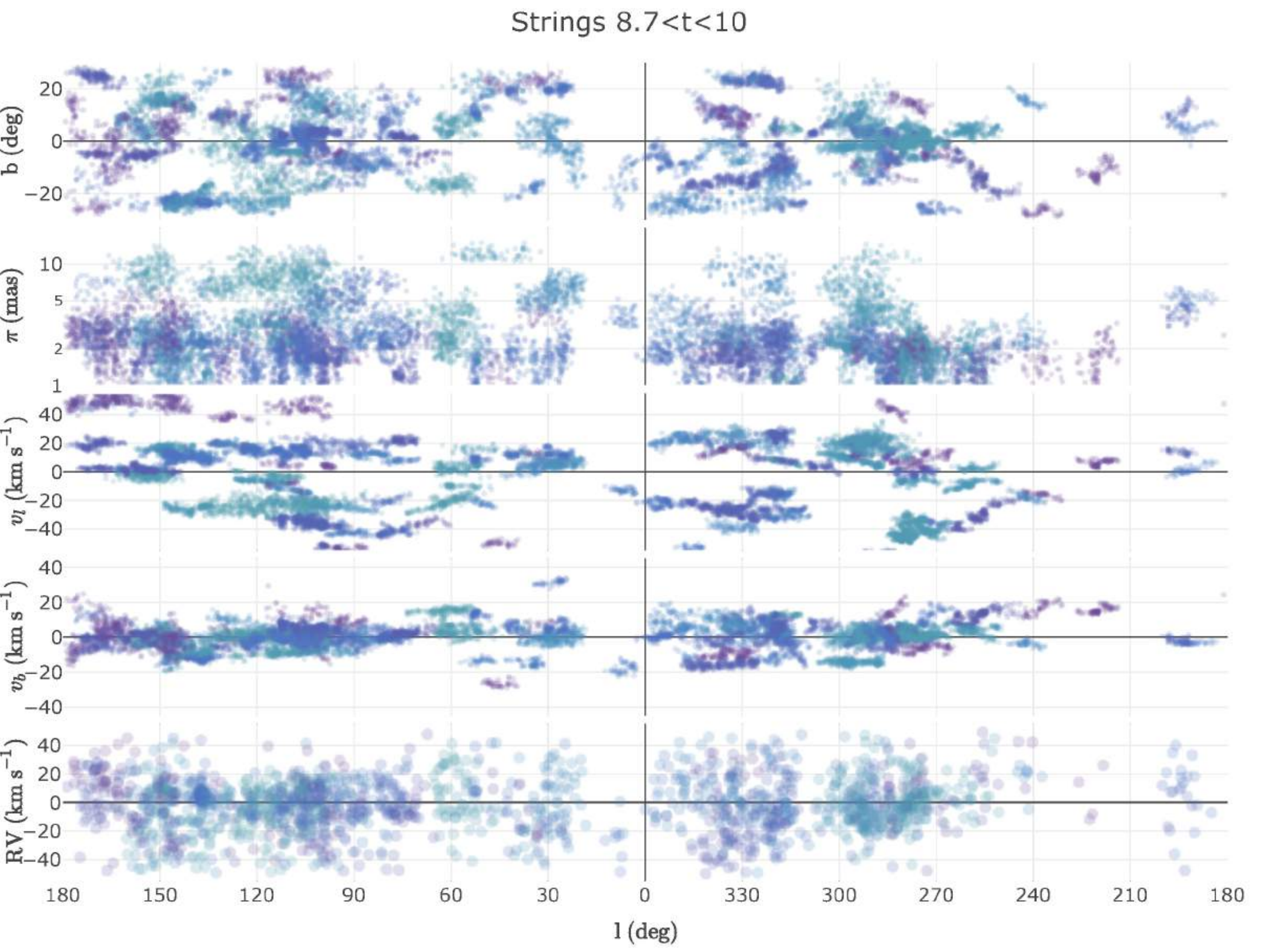}{0.5\textwidth}{}
             \fig{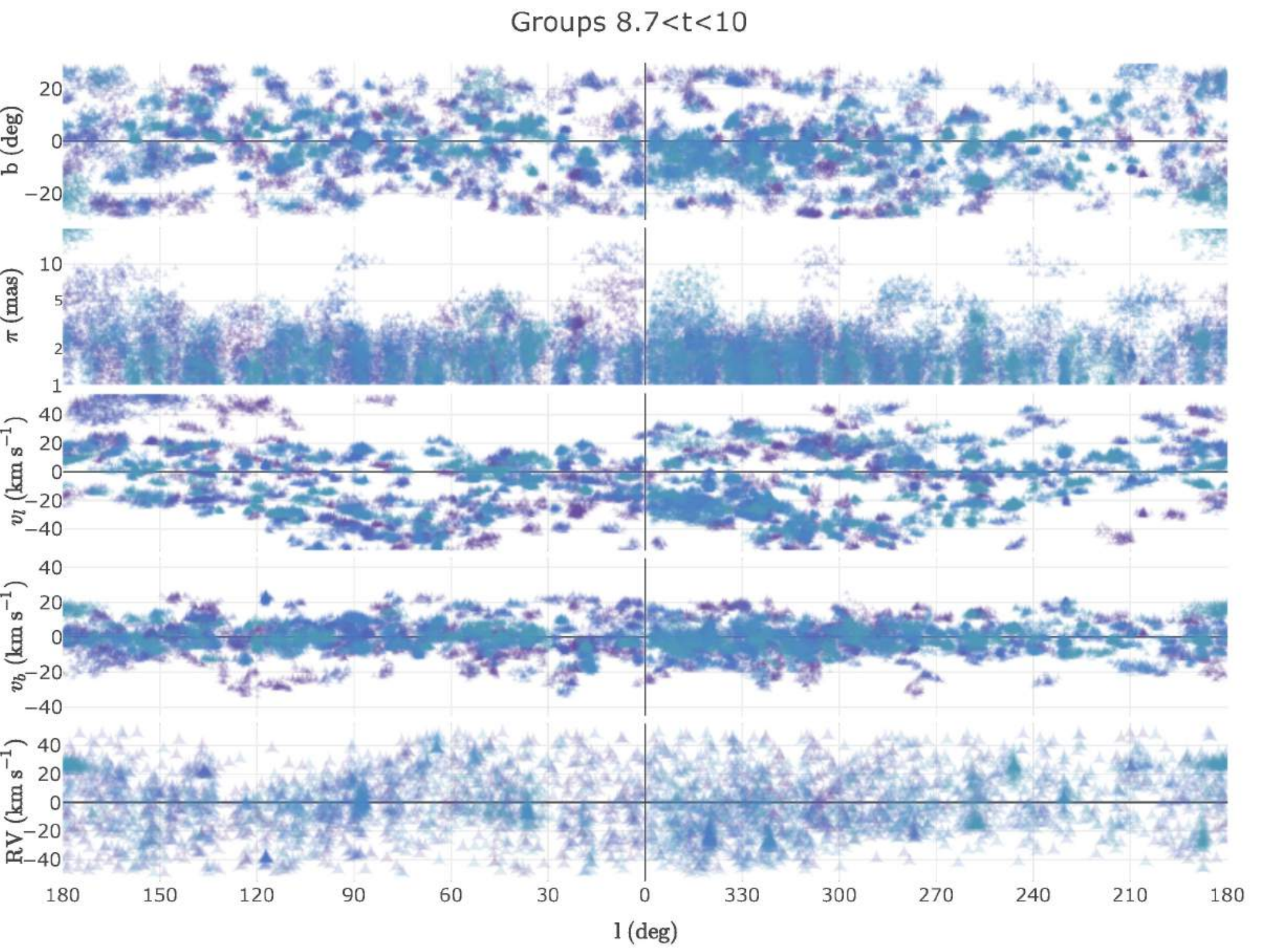}{0.5\textwidth}{}
        }\vspace{-1 cm}
        \gridline{\fig{bar.pdf}{0.6\textwidth}{}}\vspace{-0.75 cm}
\caption{Projections of the phase space of the identified strings (left) and clusters (right) in three different age ranges, with the color scale corresponding to the age from red (youngest) to purple (oldest). Velocities are given in the lsr reference frame. The separation between the structures is usually most apparent in $l$ vs. $v_l$ panel. An interactive version of the figure is available in the online version (Temporarily at \url{http://mkounkel.com/mw3d/mw2d.html}).
\label{fig:2d}}
\end{figure*}

\subsection{Properties of strings}
\begin{figure}
\epsscale{1.1}
 \centering
		\gridline{\fig{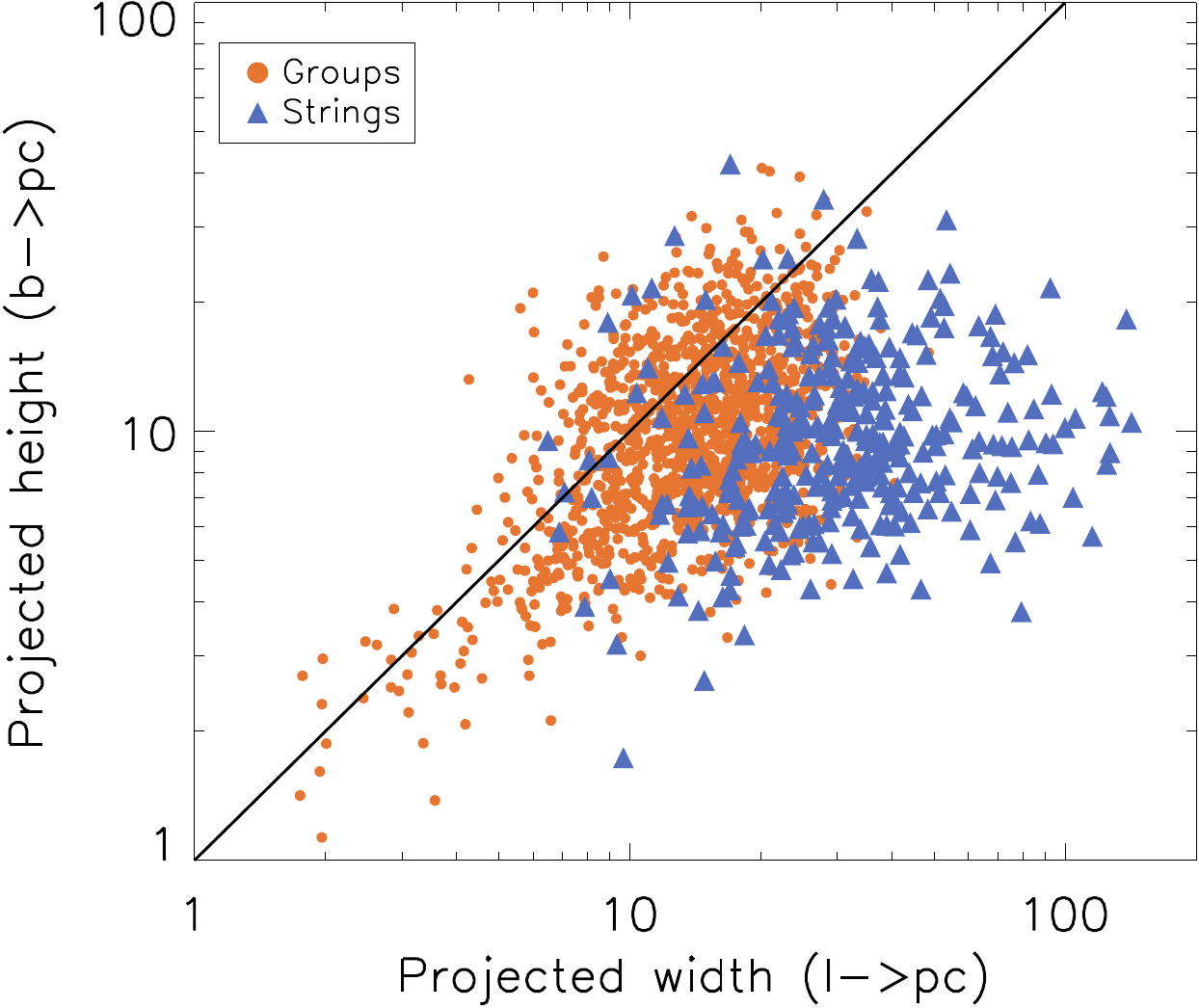}{0.23\textwidth}{}
             \fig{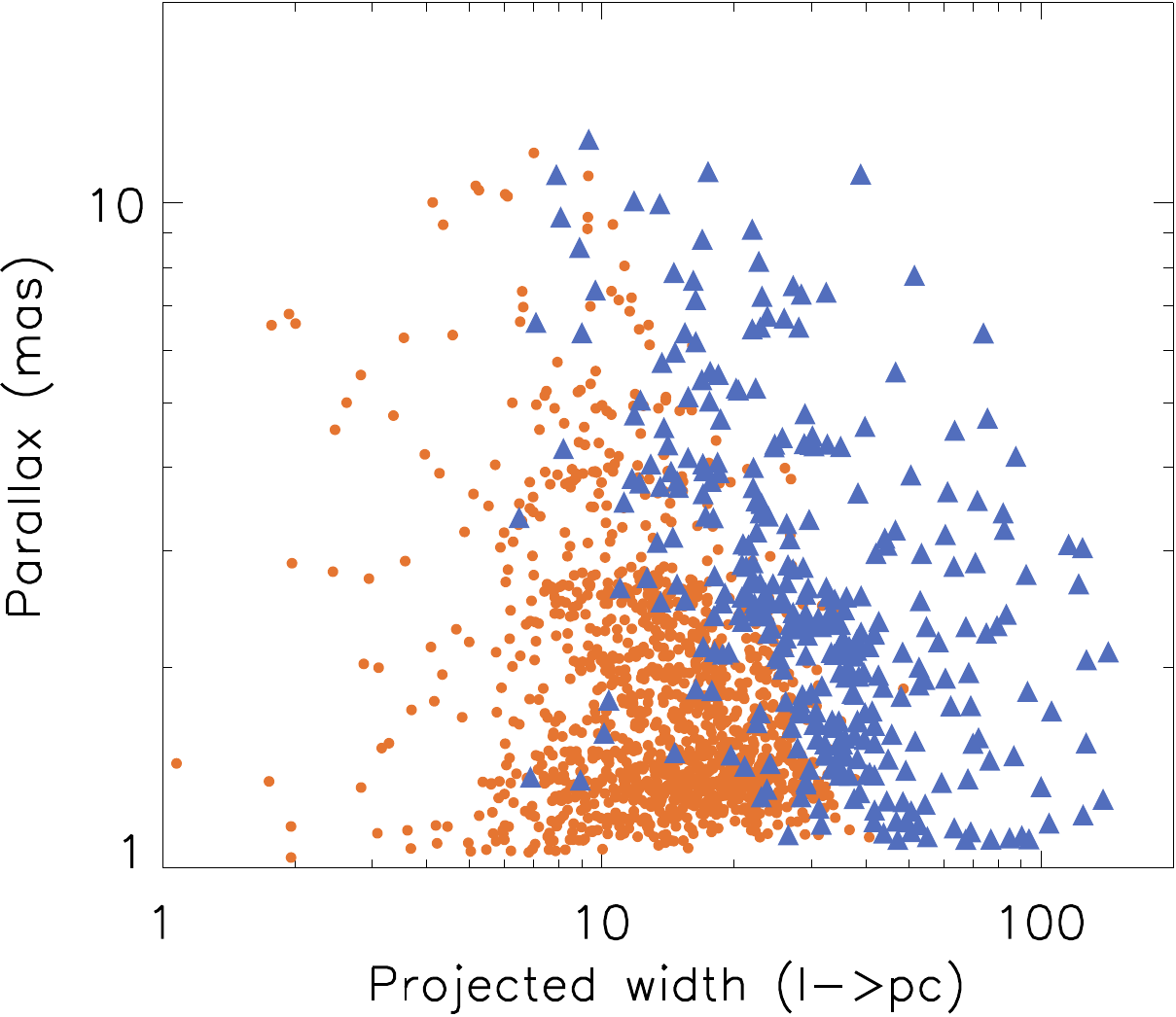}{0.23\textwidth}{}
             }\vspace{-0.75 cm}\gridline{
             \fig{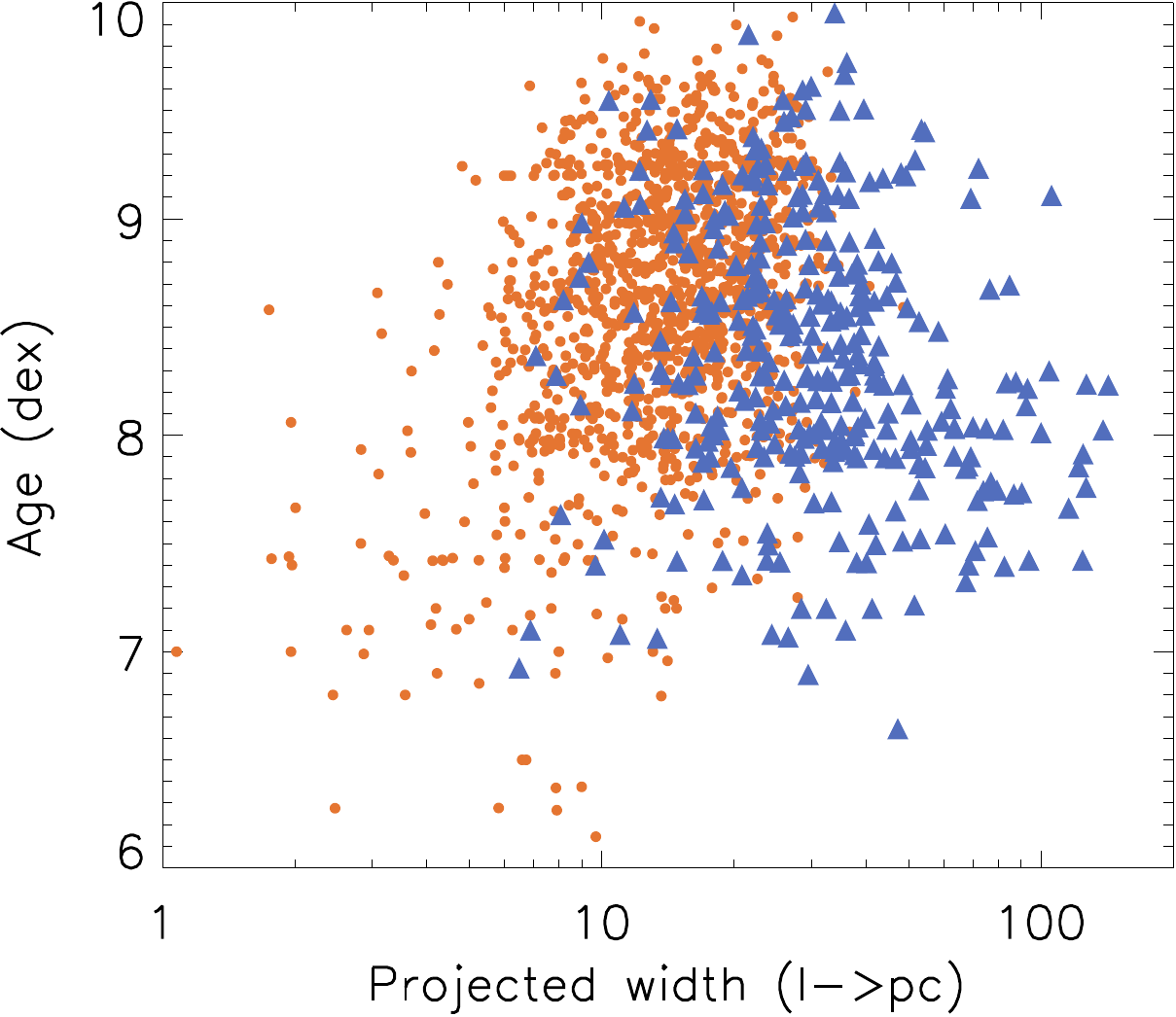}{0.23\textwidth}{}
             \fig{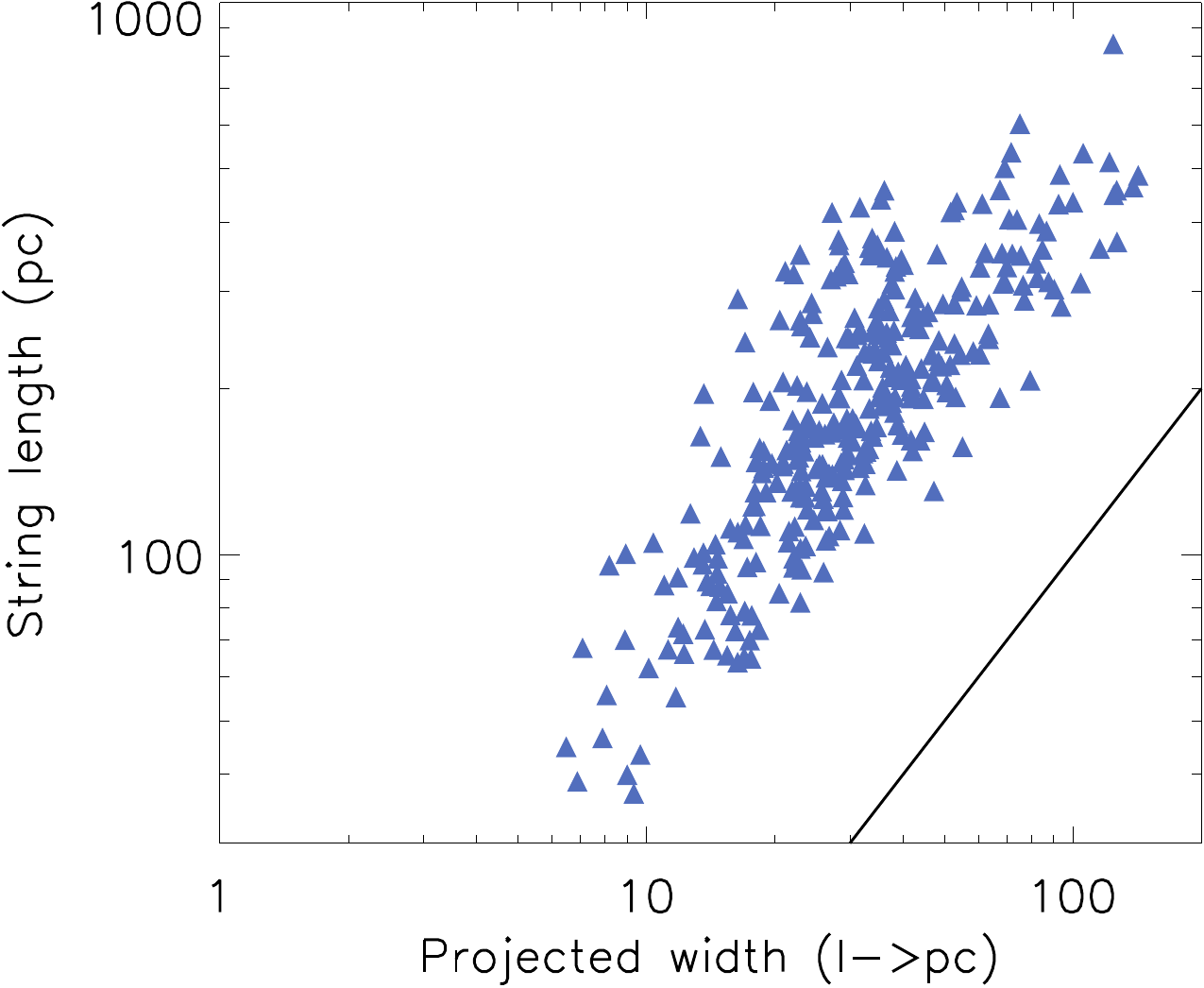}{0.23\textwidth}{}
        }\vspace{-0.75 cm}
\caption{Apparent sizes of the groups and the strings shown as a function of their relation of various dimensions, . Projected width and length are measured from the standard deviation in $l$ and $b$ converted to the physical units at the typical distance to the structure. The length is traced along the spine of the string end-to-end in the XYZ plane. Black line shows the line of equity.
\label{fig:len}}
\end{figure}

In total, 138,075 sources (approximately half of the full catalog) are found in 328 strings (Figure \ref{fig:2d}). The remaining 150,295 sources are located in 1312 groups\footnote{Include gravitationaly bound clusters, associations, and co-moving groups. Because there is no explicit definition that can currently be easily applied to separate between them observationally, the neutral term `group' will be used throughout the work. However, the issue of nomenclature, or a determination of whether there is indeed a physical difference between these concepts needs to be revisited in the future.} (Figure \ref{fig:2d}). To the best effort, the separation into groups and strings was done self-consistently through the visual examination, however, in some cases the difference between the two can be poorly defined, as most groups do show some degree of extension along $l$ (Figure \ref{fig:len}). There is some bias in the identification ability as a function of distance, as nearby short strings would be more easily apparent due to their large angular size, compared to those that are further away. The typical length of individual strings when traced along their full extent along their spine is $\sim200$ pc.

\begin{figure}
\epsscale{1.1}
\plotone{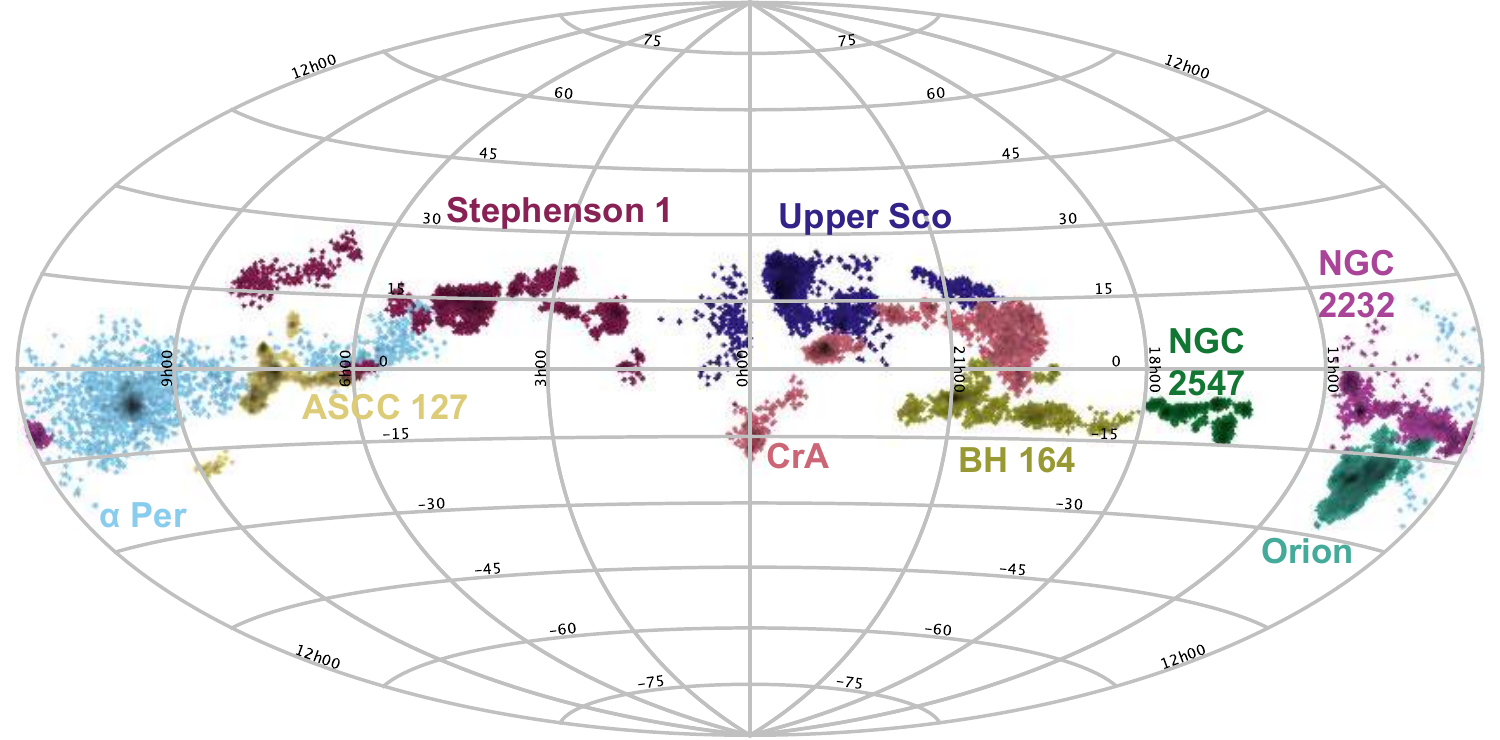}
\caption{Some examples of notable structures, plotted in the galactic coordinates. See the interactive version of Figure \ref{fig:2d} for full projections of all of the individual strings.
\label{fig:example}}
\end{figure}

Nearby strings in particular appear to span large angular sizes. While it is not physically the largest, one of the most spectacular ones counts the cluster $\alpha$ Per as a part of it, and it extends for over 120$^\circ$ (Figure \ref{fig:example}). Another peculiar one is located just north of the Orion Complex, containing NGC 2232, and it is probably the parental population that has formed Betelgeuse. Orion Complex itself can be imagined as a single string-like structure, although it has a more complex structure and star forming history than what is typical. Sco-Cen OB association appears to consist of two closely connected strings: one that extends from Upper Sco to Lower Centaurus-Crux (as is typically recognized), and another that extends from Lower Centaurus-Crux to Corona Australis star forming region, forming a shape of a wishbone that diverges both spatially and kinematically on one end. Although they are related, they were treated separately during the selection. However, both of them are physically smaller than a typical string.

Recently a similar extended structure has been identified in the solar neighborhood, which is referred to as Pisces-Eridanus stream \citep{meingast2019,curtis2019}. It is located at a distance of $\sim$ 100 pc, spans $\sim120^\circ$, and has an age of $\sim$120 Myr. We do not recover it due to its height above the galactic plane. However, it is likely to be similar to the strings we do find.

Looking at the distribution of all the sources as a whole, kinematically, there does not appear to be a significant difference between strings and groups, both showing a similar structure in the position-velocity diagrams. However, there is a significant difference in the age distribution between the two (Figure \ref{fig:age}). Almost all of the newly formed stars can be collected into various strings. And while some strings may survive up to a few Gyr, they become significantly less numerous at ages beyond 8.5 dex compared to the younger counterparts. On the other hand, the age distribution of non-strung together groups does not peak until 8.6 dex.

\begin{figure}
\epsscale{1.1}
 \centering
		\gridline{\fig{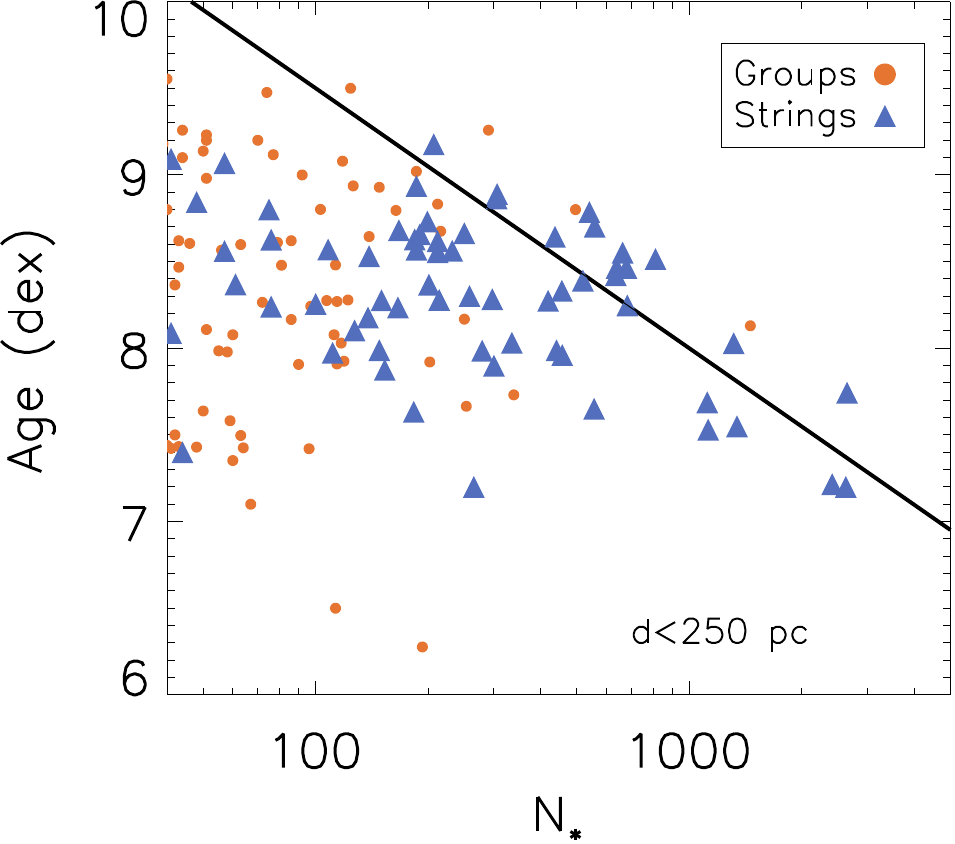}{0.23\textwidth}{}
             \fig{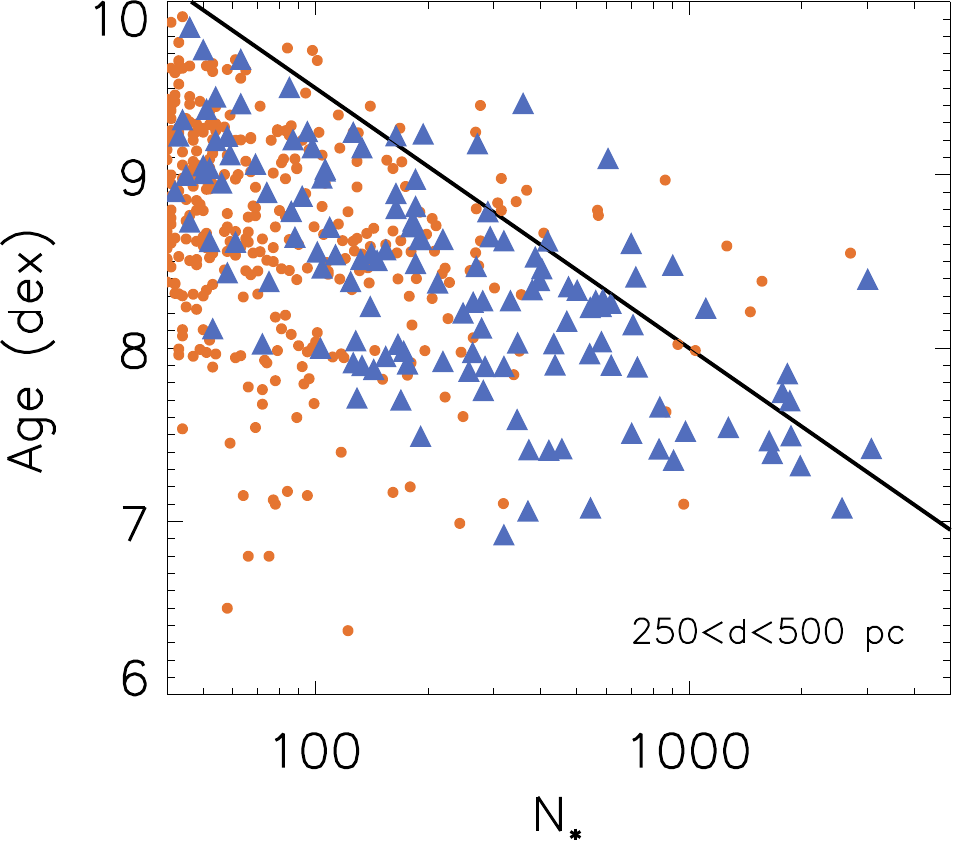}{0.23\textwidth}{}
        }\vspace{-0.75 cm}
		\gridline{\fig{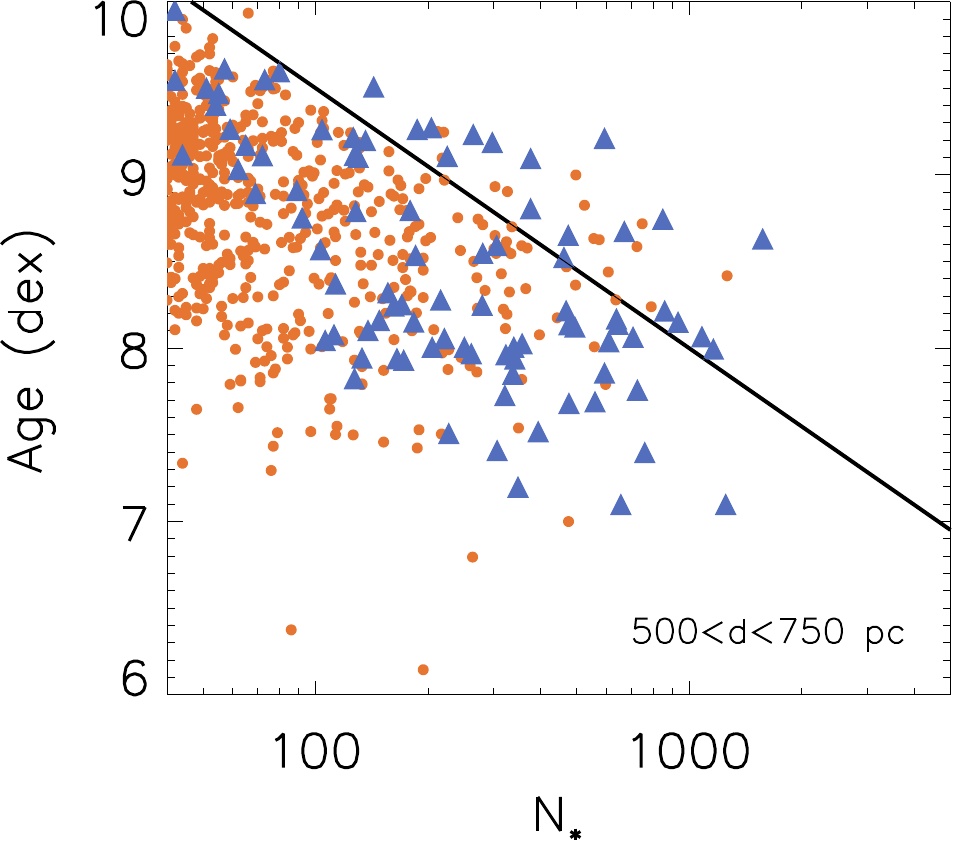}{0.23\textwidth}{}
             \fig{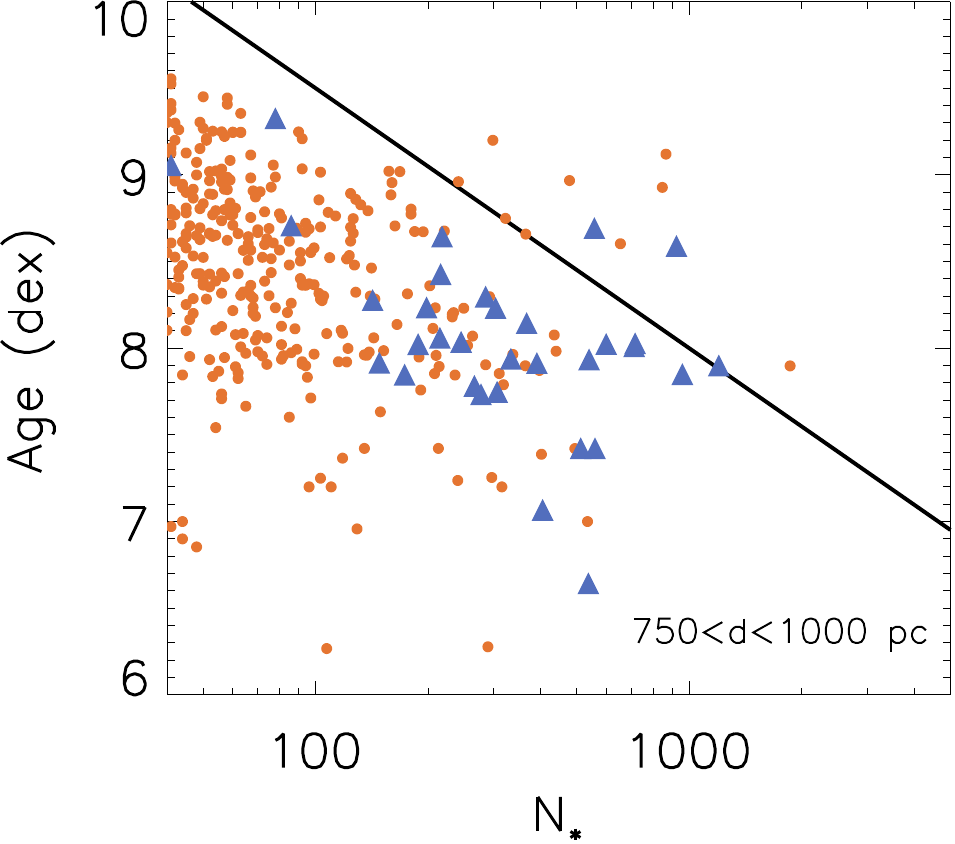}{0.23\textwidth}{}
        }\vspace{-0.75 cm}
\caption{Number of stars in individual groups and strings in various distance slices. The black line shows an approximate evolution of the group size is the power law with the slope $\sim-1.5$.
\label{fig:nstar}}
\end{figure}

While there is no correlation between the size of a structure and its age (Figure \ref{fig:len}), the number of sources in individual groups and strings appears to decrease strongly for older structures. There is no characteristic size of the newly born populations, as they can contain anywhere from a few dozen to several thousand stars. On the other hand, oldest populations rarely contain more than 100 stars. This effect is not caused by the distance-driven incompleteness (Figure \ref{fig:nstar}). Assuming that it is possible to map the most massive young populations to the most massive old ones, the number of surviving sources within a structure can be roughly characterized as a function of age with a power law $N_*= N_\circ\times10^{-(t-t_\circ)/1.5}$. Regarding the apparent sizes, although there may be some apparent decrease in the projected size in age in Figure \ref{fig:len}, it is largely driven by the number of stars in each population; there is no trend in age with the absolute length that is traced in 3d end to end.

\begin{figure}
\epsscale{1.1}
 \centering
		\gridline{\fig{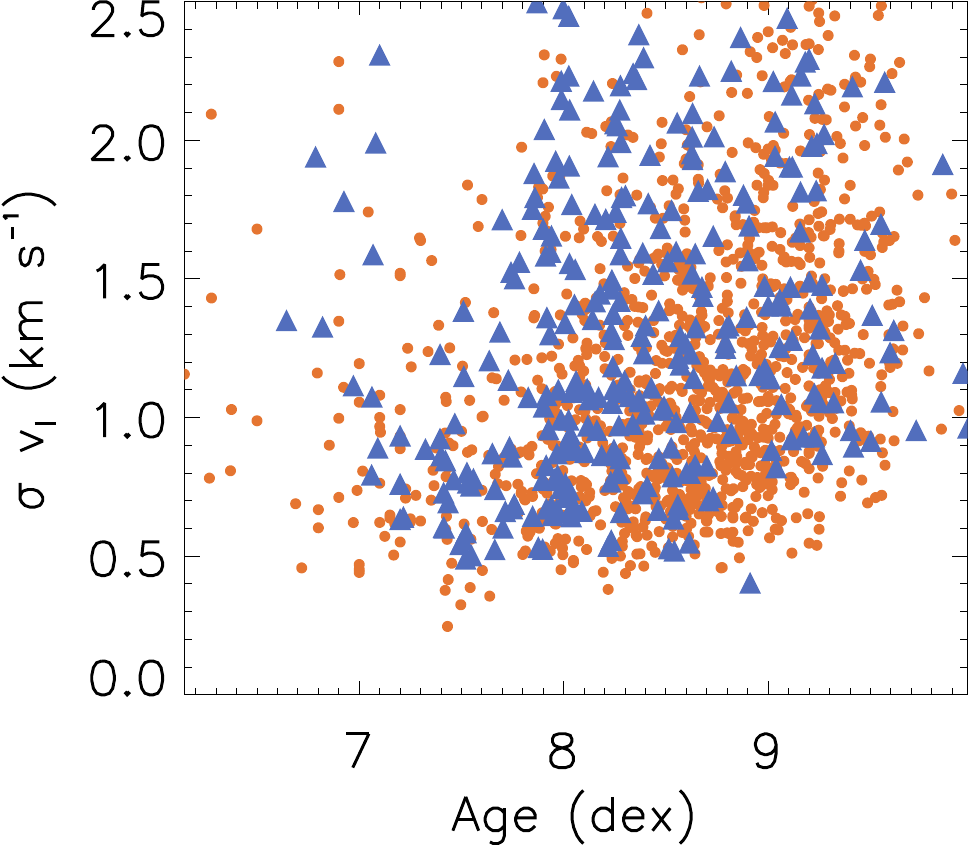}{0.23\textwidth}{}
             \fig{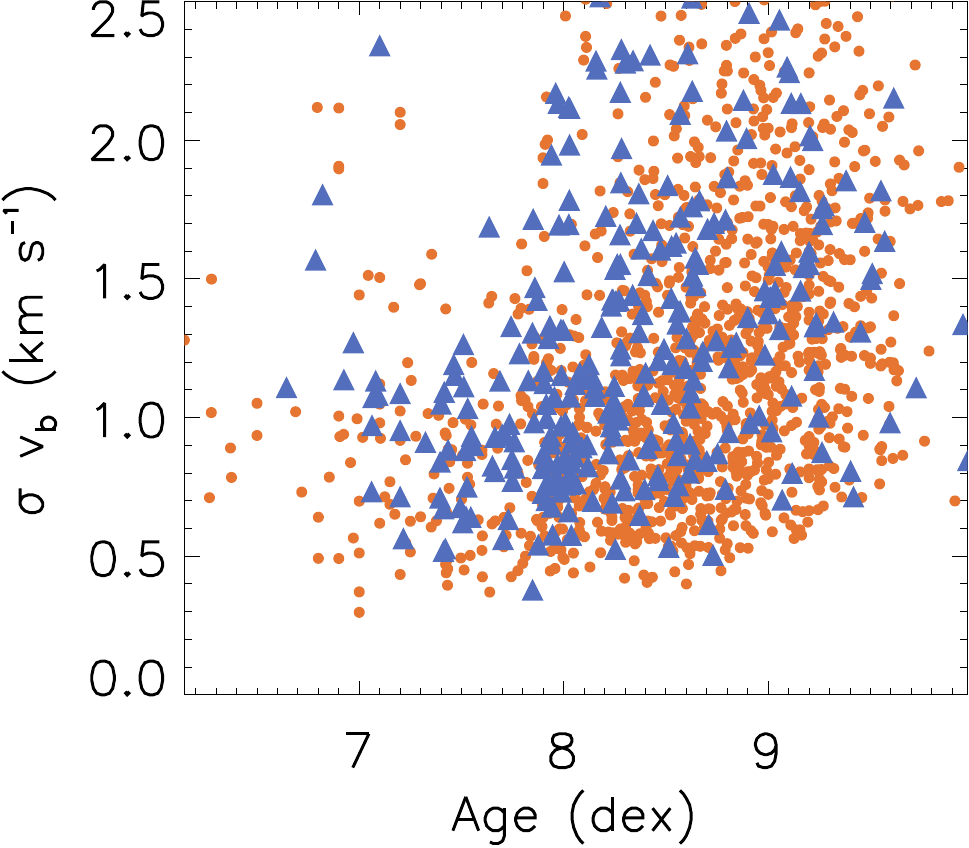}{0.23\textwidth}{}
             }\vspace{-0.75 cm}\gridline{
             \fig{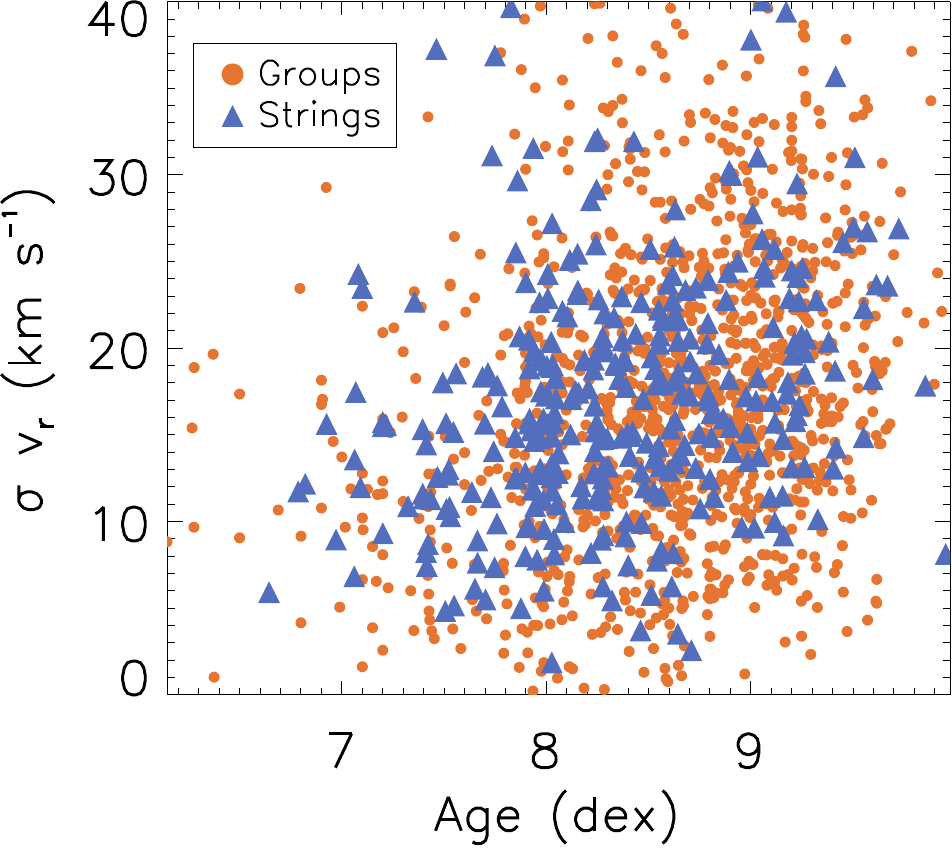}{0.23\textwidth}{}
        }\vspace{-0.75 cm}
\caption{Velocity dispersion in 3 dimensions ($l$, $b$, and radial) for individual groups and strings. The underlying spatial dependence in kinematics as a function of $l$ was subtracted prior to measuring the dispersion.
\label{fig:dv}}
\end{figure}

Velocity dispersion for the surviving members of the individual structures that have not been dissolved into the field changes as a function of age as well, with $\sigma_{v_l}$ and $\sigma_{v_b}$ increasing from $\sim$0.75 \kms\ to $\sim$1.5 \kms\ between ages of 7 and 9 dex. Because radial velocity measurements from \textit{Gaia} DR2 have large uncertainties, and they have not been examined for multiplicity, the specific $\sigma$ measurements are less meaningful, but they do show a similar underlying trend. The measured velocity dispersions are the lower limit for the populations, if it were possible to incorporate all the original members that have long since dissolved into the field. However, it should be noted that populations that form with a smaller number of stars tend to have a smaller velocity dispersion than those that are more massive at any given age, therefore the evolution towards higher velocity dispersion is in part due to a relaxation of the smaller groups over time. However, if it were possible to account for all the members that have formed within any given population and have since dissolved into the field, the absolute velocity dispersion should increase more significantly.

In addition to the strong prominence of stringing in younger populations, many strings do not appear to have an open cluster at their core. Because of this, it is unlikely their shape originates from tidal stretching in comparison to the various streams associated with globular clusters or dwarf galaxies. Rather, for young string this extended shape is likely primordial, mirroring the shape of the molecular cloud from which they have formed. Possibly, some of them originate from giant molecular filaments \citep[e.g.,][]{ragan2014,zucker2018a}. While it may be surprising that they can survive as quasi-coherent structures into the Gyr age, these strings do not show a significant evolution beyond slowly dissolving into the field, and the oldest strings are most likely the remnants of some of the most massive structures that have originally formed. While we characterize their present day structure and kinematics in the next section, we defer a deeper exploration of various formation scenarios and tests of dynamical stability in the Galactic potential to future work.

\subsection{Large scale structure and dynamics}

\begin{figure*}
\epsscale{1.1}
 \centering
		\gridline{
             \fig{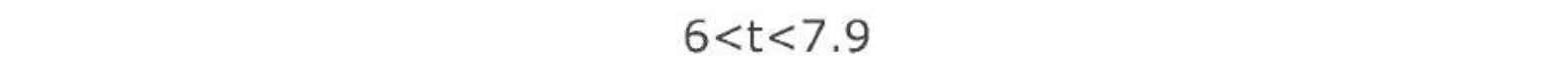}{0.5\textwidth}{}
             \fig{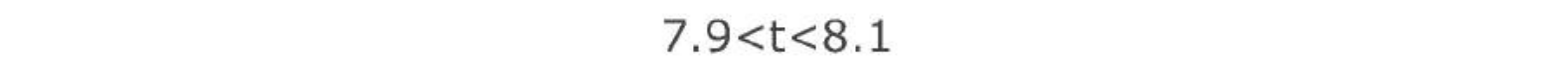}{0.5\textwidth}{}
        }\vspace{-1 cm}\gridline{
		\fig{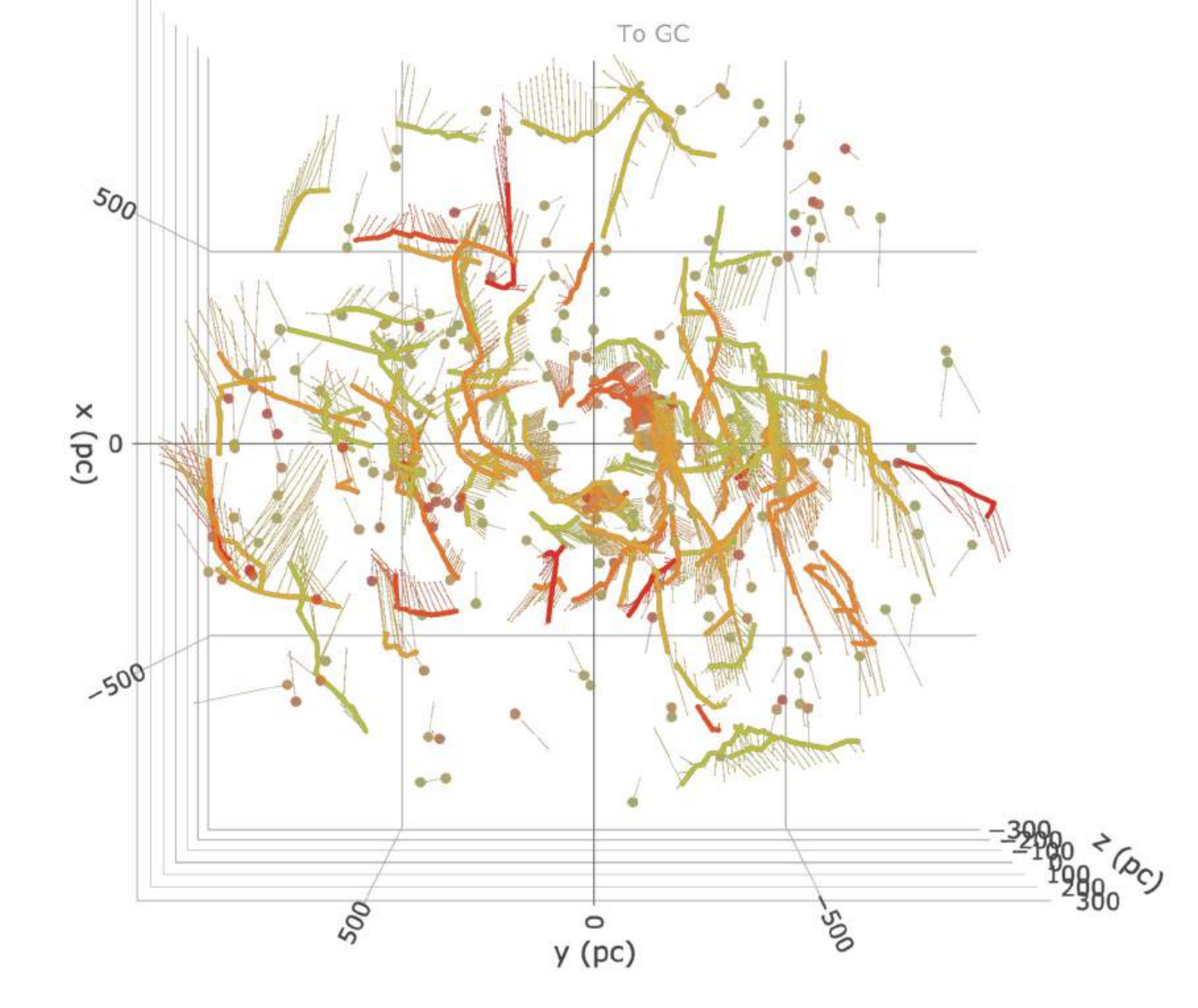}{0.45\textwidth}{}
             \fig{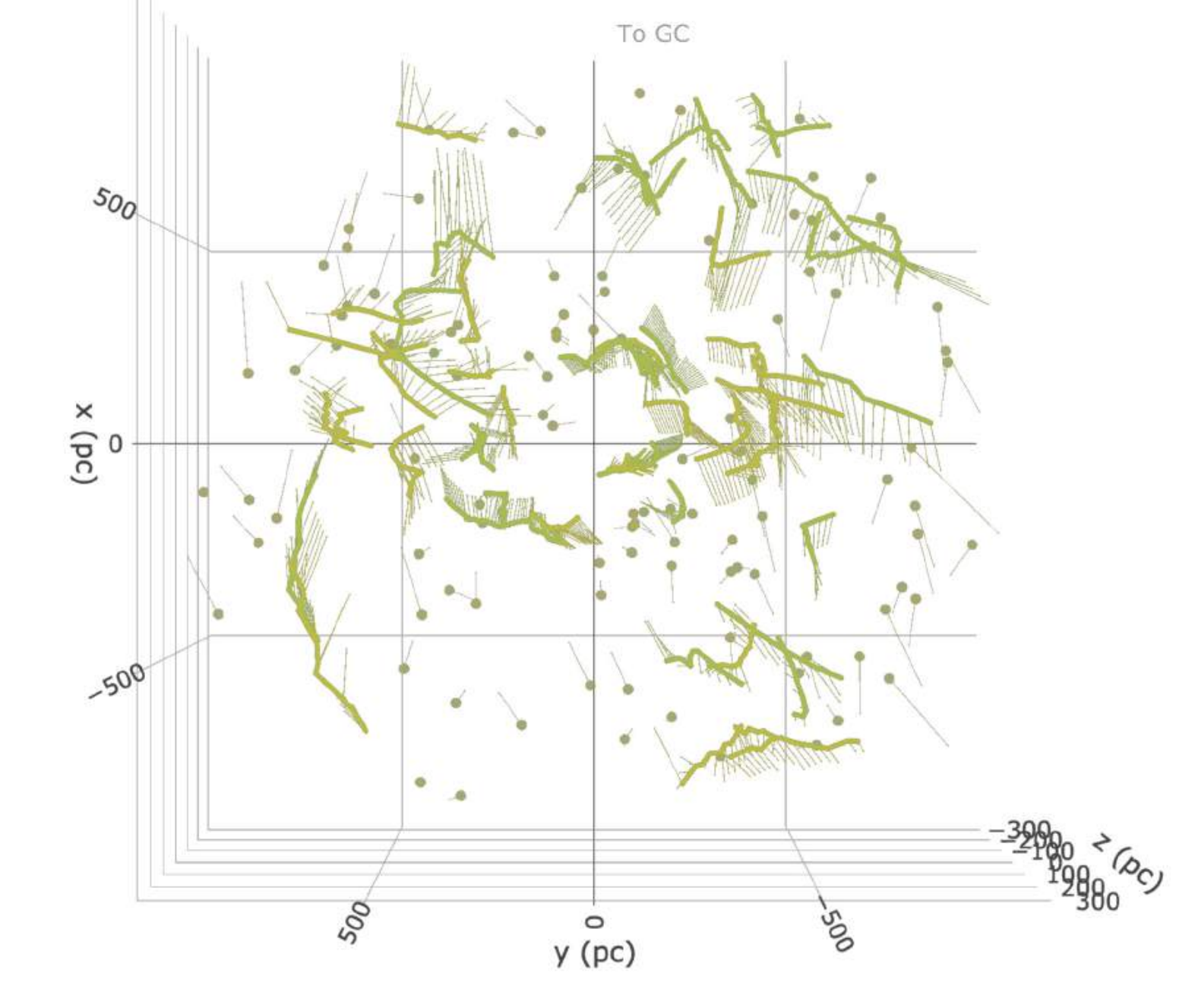}{0.45\textwidth}{}
             }\vspace{-1 cm}\gridline{
             \fig{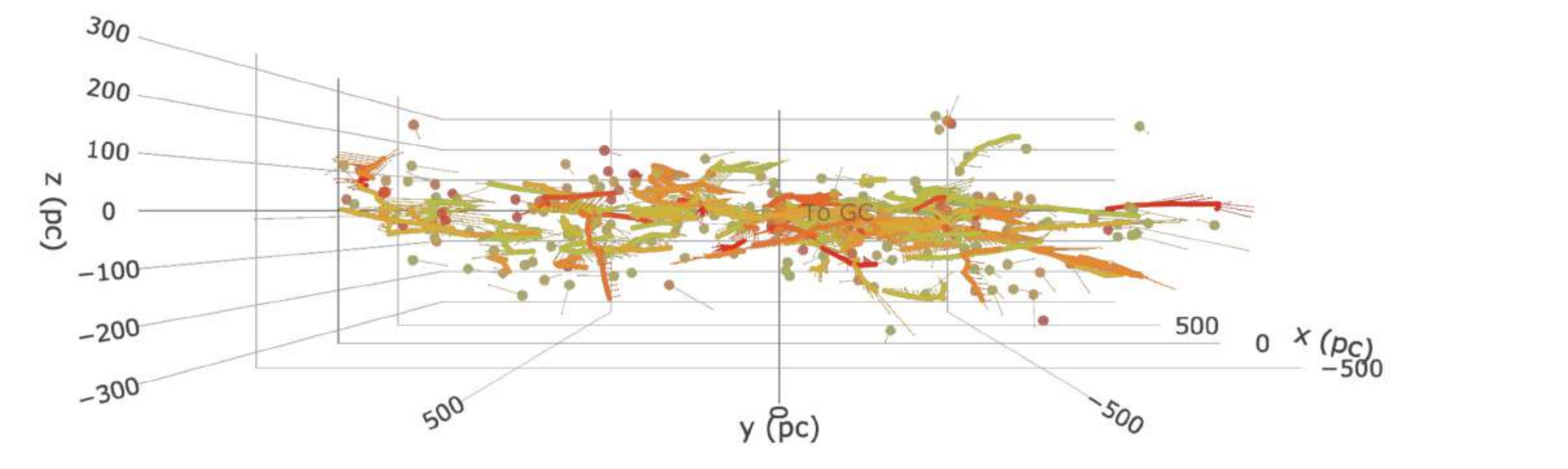}{0.5\textwidth}{}
             \fig{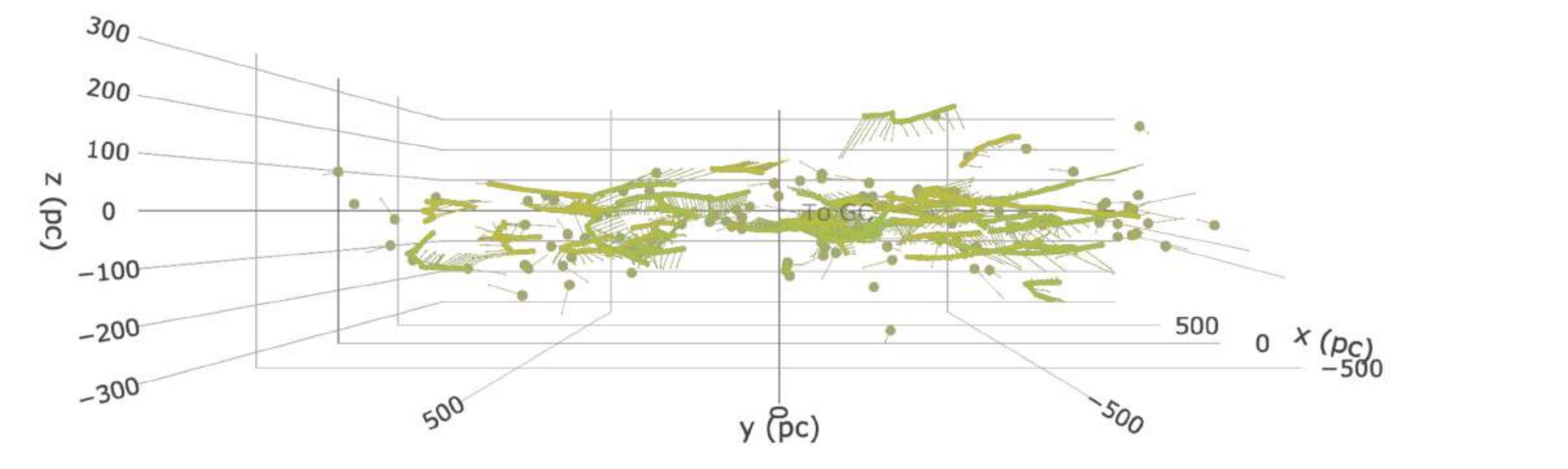}{0.5\textwidth}{}
        }\vspace{-0.75 cm}\gridline{
             \fig{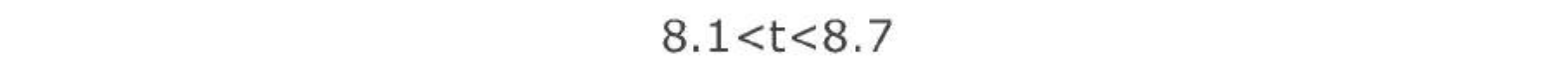}{0.5\textwidth}{}
             \fig{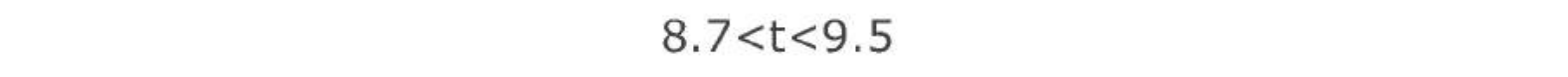}{0.5\textwidth}{}
        }\vspace{-1 cm}\gridline{
             \fig{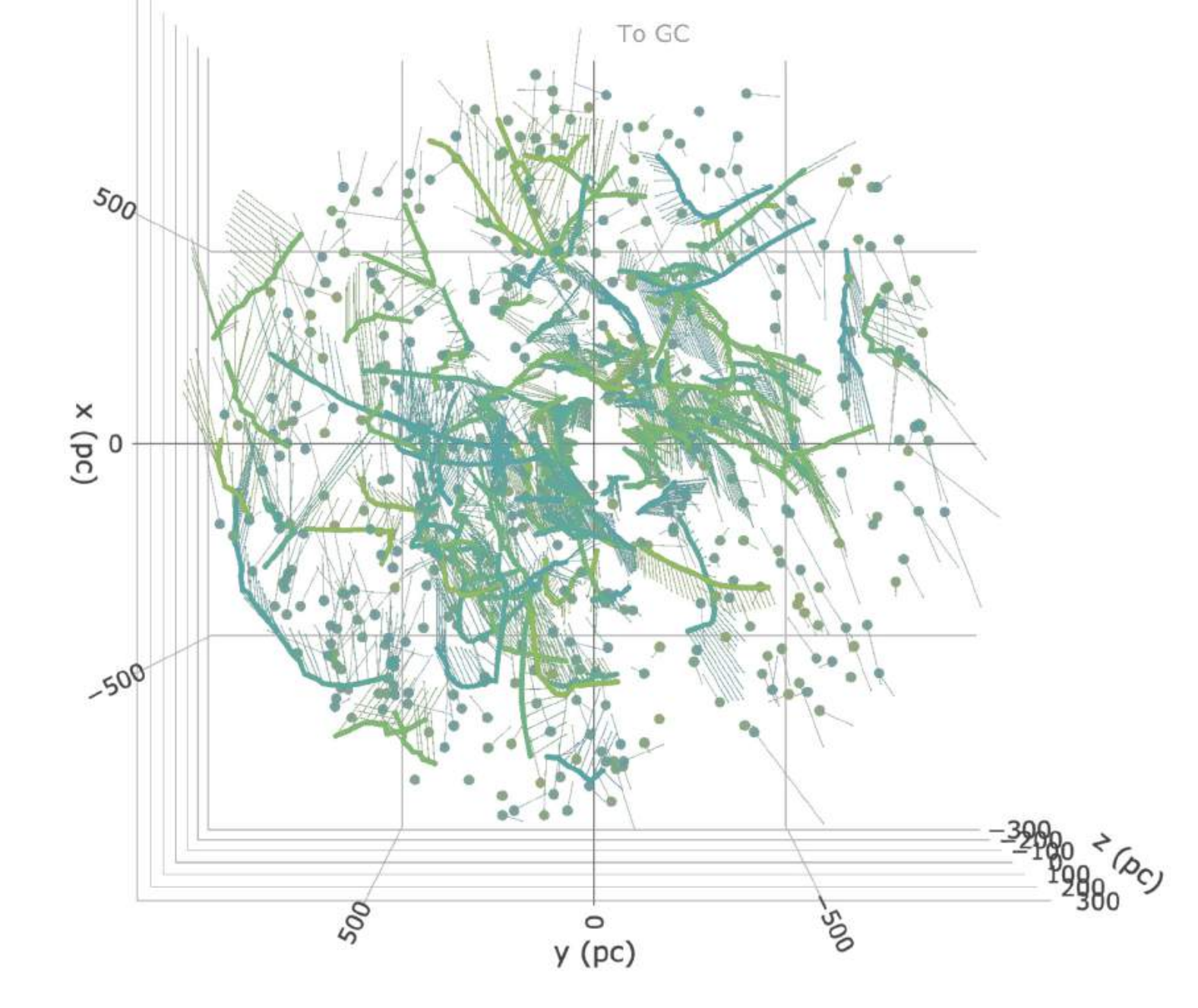}{0.45\textwidth}{}
             \fig{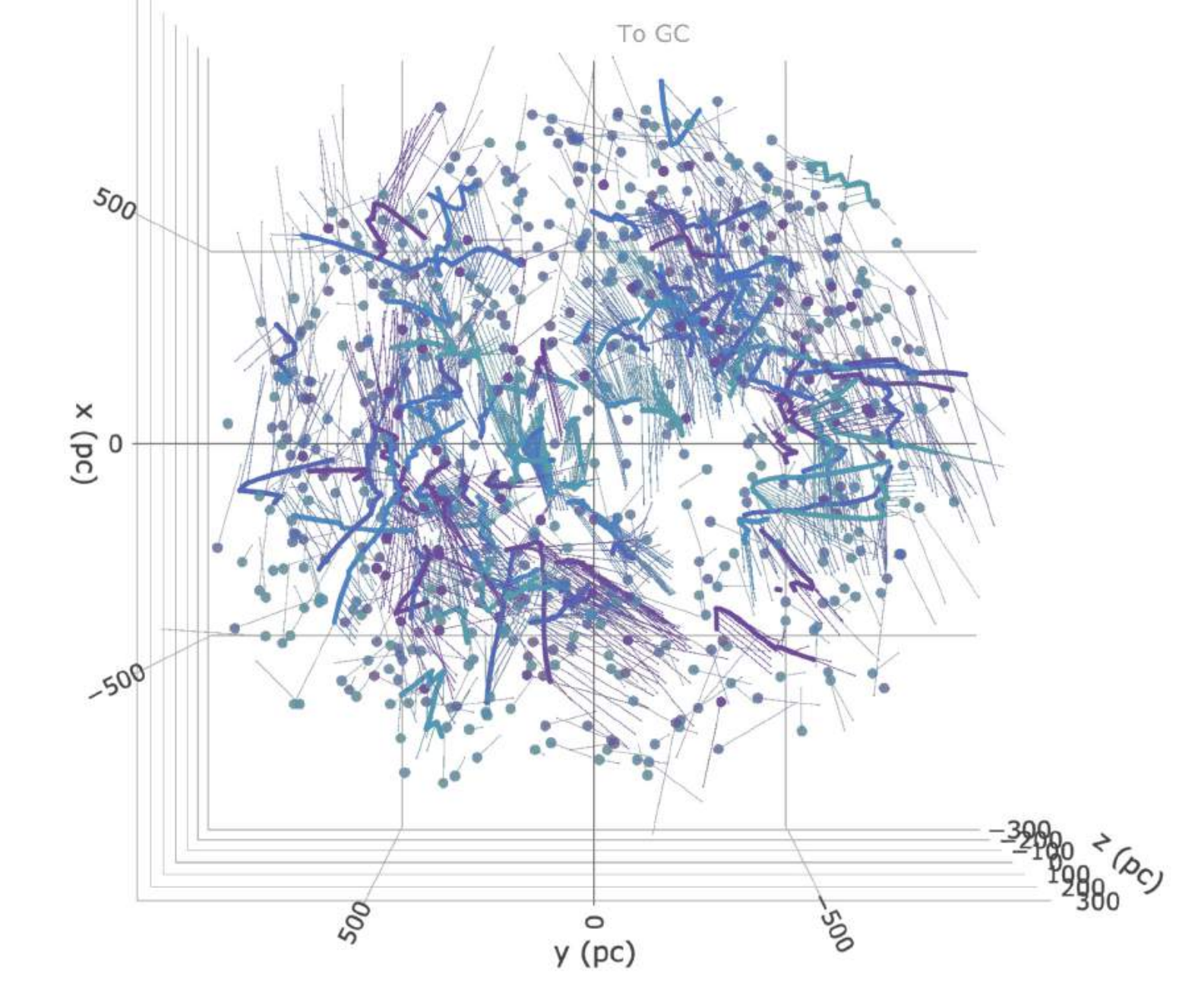}{0.45\textwidth}{}
        }\vspace{-1 cm}\gridline{
             \fig{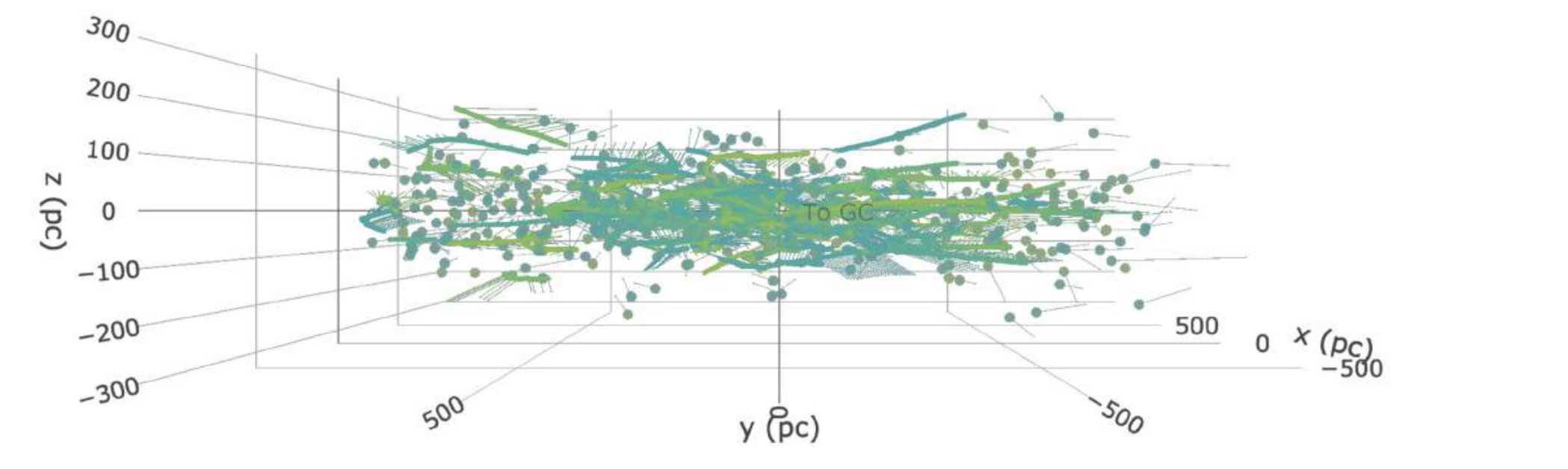}{0.5\textwidth}{}
             \fig{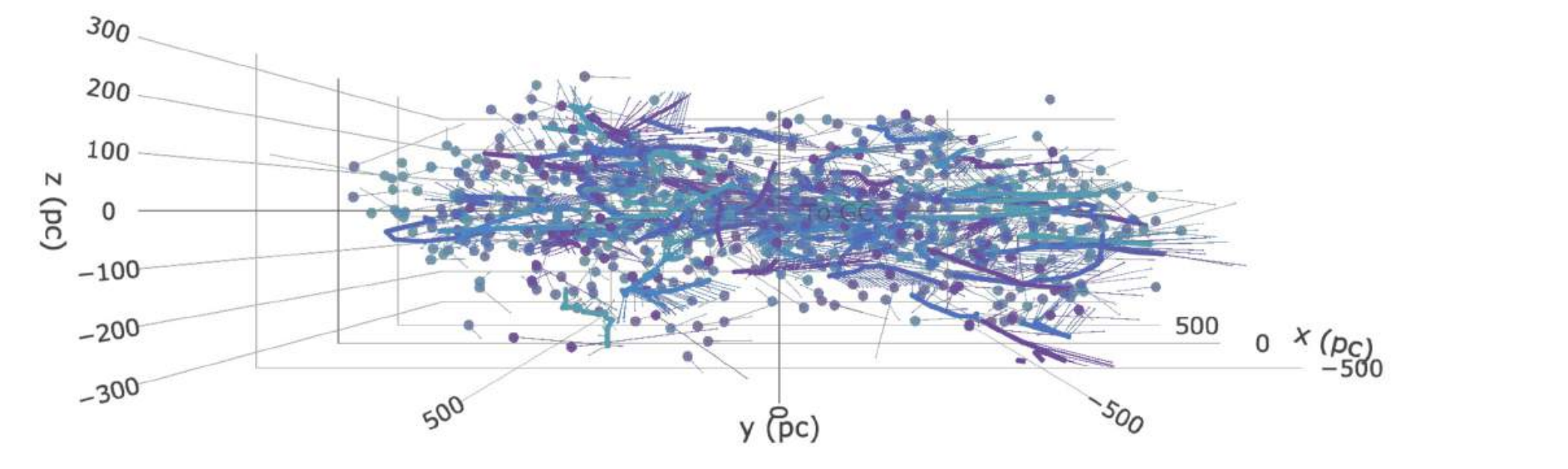}{0.5\textwidth}{}
        }
        \vspace{-0.75 cm}
        \gridline{\fig{bar.pdf}{0.6\textwidth}{}}\vspace{-0.75 cm}
\caption{Face-on and edge-on map of the solar neighborhood, with the identified strings and groups, color-coded by age, in various age slices. Thinner lines show the velocities of the structures over the next 5 Myr, ignoring the galactic potential. An interactive 3d plot is available in the online version. (Temporarily at \url{http://mkounkel.com/mw3d/mw3d.html)}.
\label{fig:3d}}
\end{figure*}

To place the individual strings into the physical galactic reference frame, we first determine their individual 2d structure. To trace the spine of each string, we average $b$, $\pi$, and the kinematics in 1 degree bins along $l$, and then smooth the resulting arrays with the Savitzky-Golay filter (Table \ref{tab:string}) using a $20^\circ$ kernel or half the full width of the string in $l$, whichever one is the smallest. This is done to avoid strong fluctuations in the averages that could result from too few sources in a given bin. The positions are then converted into the heliocentric XYZ reference frame, and the UVW velocity vectors are computed using GalPy \citep{galpy}.

Strings with ages $<7.9$ dex appear to jointly form a coherent structure, Stream 1, that is $\sim$300 pc wide, extending beyond the boundary probed by this sample (Figure \ref{fig:3d}). However, older strings do not coincide with it. Instead, those with ages between 8.1 and 8.7, appear to form a separate coherent Stream 2 that is oriented to the younger one by $\sim60^\circ$. Furthermore at ages beyond 8.7, two other streams (one that is located at $Y\sim500$ pc, Stream 3 - which has also recently been identified as a stellar overdensity by \citet{miyachi2019}, and one that is at $Y\sim-700$ pc, Stream 4) may be apparent. In all of the cases, the strings appear to be preferentially oriented perpendicularly to these streams.

\begin{figure*}
\epsscale{1.1}
\plotone{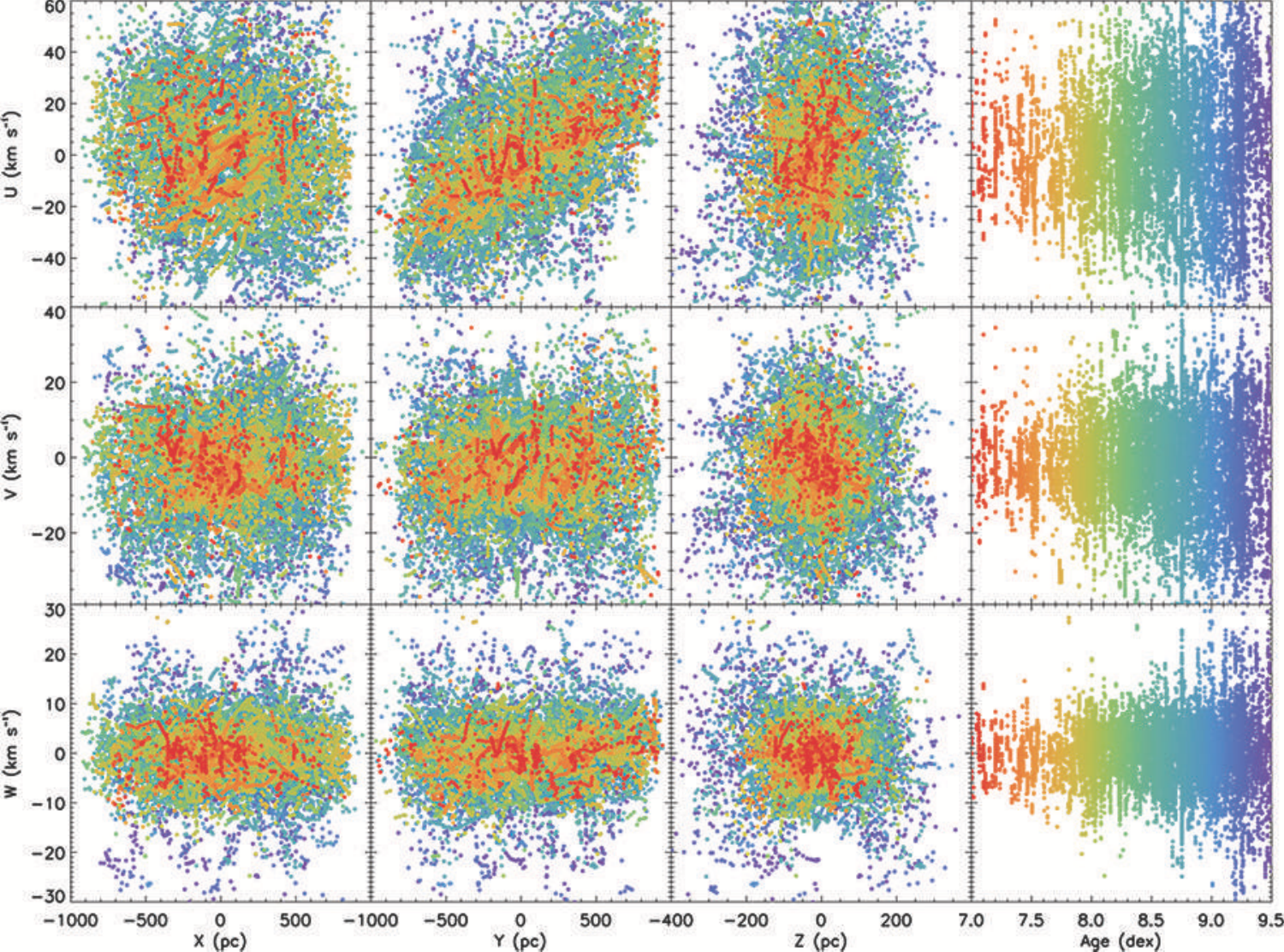}
\plottwo{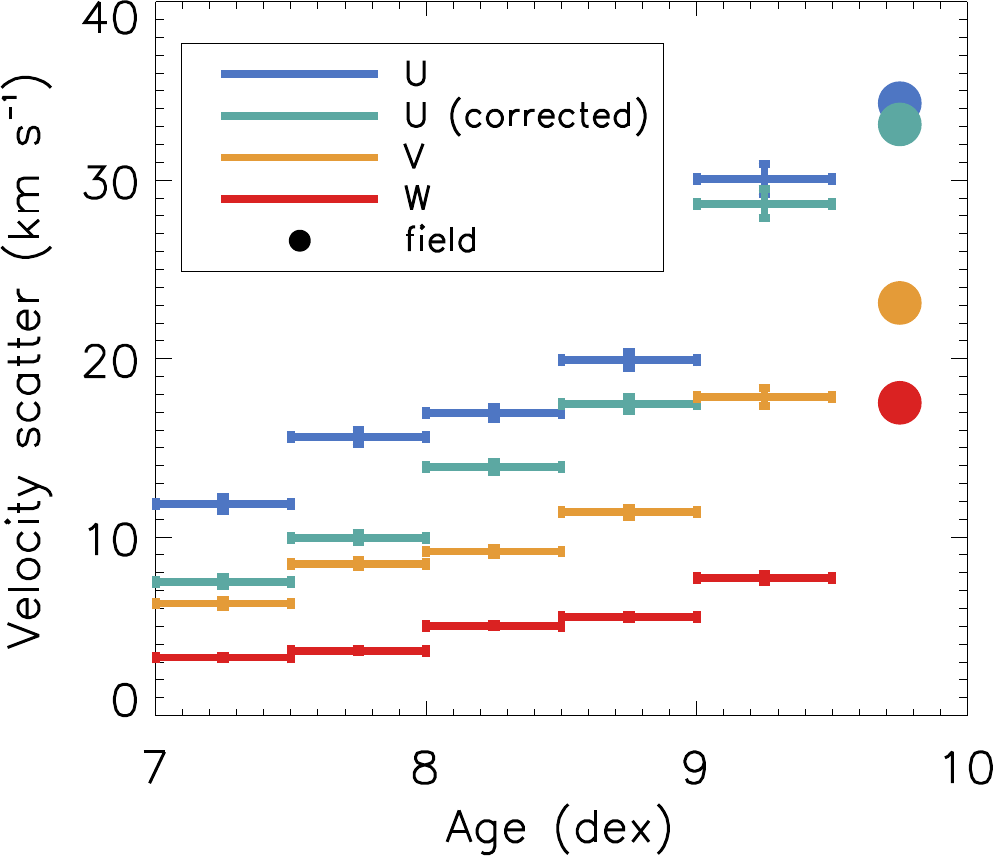}{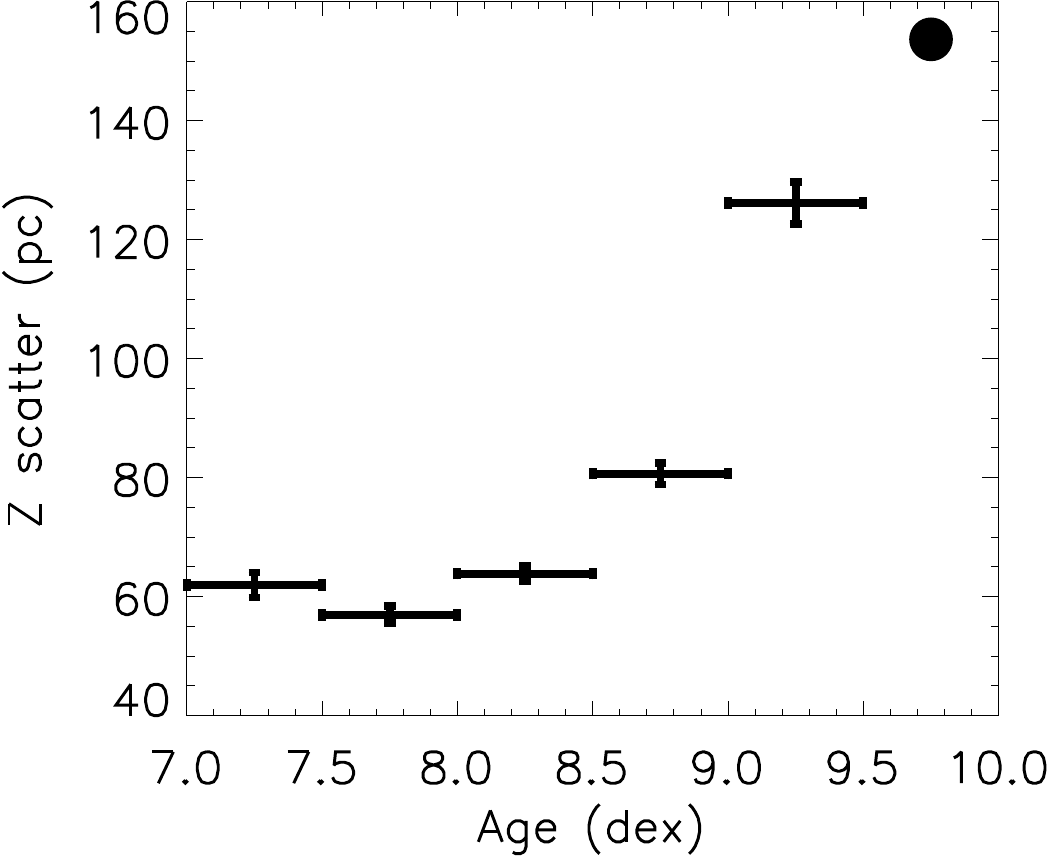}
\caption{Top: Correlations between position and velocity as a function of age in the solar neighborhood, relative to lsr. Bottom: Scatter of the velocities in rectangular coordinates (left) and the height of the galactic disk (right) of the clustered structures as a function of age. Both the measured $U$ and the $U$ corrected for the cross term in $Y$ specified by the Equation \ref{eqn:corr} are shown. The circle in the last bin shows the scatter of all of the unclustered sources.
\label{fig:vxyz}}
\end{figure*}

Kinematically, the streams are, on average, in the local standard of rest (Figure \ref{fig:vxyz}). However, while the lsr corrections from \citet{schonrich2010} hold on average, there are unaccounted cross terms. While $V_{ave}\sim V_\odot$, and $W_{ave}\sim W_\odot$, there is a strong dependence between U ($\equiv\Delta$X) and Y. This dependence can be characterized by 

\begin{equation}\label{eqn:corr}
U_{ave} \text{(km s}^{-1}\text{)}=U_\odot+2.14-0.030\times(Y\text{pc})
\end{equation}

Previously, \citet{schonrich2012} has observed a signature of galactic rotation in the combination of the stellar velocities. Upon examining unclustered sources within 1 kpc, two cross terms are apparent: a strong correlation between $Y$ and $U$, as well as one between $X$ and $V$, that do indeed resemble a rotation. On the other hand, clustered sources do not show a correlation between the $X$ position and $V$ velocity, making their global motion in the same volume of space resemble a (local) shear as opposed to rotation, as can be seen in Figure \ref{fig:3d}.

The scatter in $U$, $V$, and $W$ for the clustered structures increases as a function of age (Figure \ref{fig:vxyz}), and the scatter of the remaining sources that are not part of the clustered catalog is larger yet. We can assume that most of the unclustered sources are representative of the Milky Way disk population and are older than a few Gyr. Similarly, the scatter in $Z$ also increases with age, although this is less apparent for the younger structures due to a large-scale warp throughout the Stream 1. The Gould Belt, which is formed by the nearby star-forming regions, is tilted relative to the Galactic plane by $\sim20^\circ$; several mechanisms have been proposed for its formation in the past assuming a circular shape, although recently it has been shown that Gould's Belt does not form a ring \citep{zari2018}. Instead, it is likely that its tilt is part of a larger warp of the Local Arm itself. Such a warp is not apparent in the older Streams, although due to them being thicker, it would be more difficult to confirm.

\section{Discussion}\label{sec:concl}

In this work we analyzed the distribution of sources in 5-dimensional space in \textit{Gaia} DR2 data. Using HDBSCAN we searched for extended coherent structures. We found a total of 1900 clusters and co-moving groups within 1 kpc, consisting of $\sim$300,000 sources, and we measured ages for these groups using a combination of isochrone fitting and machine learning.

Approximately half of all of the clustered sources are found in various strings that typically span $\sim$200 pc, and they are oriented in parallel to the galactic plane. In most cases we expect that these strings are primordial and are not formed as a result of a tidal disruption of the clusters. This stringing is most prominent in the youngest populations. However some may survive into Gyr age, dissolving the majority of their sources into the field, and increasing in their velocity dispersion.

The distribution of strings varies significantly as a function of age. Strings of similar ages (to within a few hundred Myr) appear to form four coherent streams, and they are preferentially oriented orthogonal to them. 

Stream 1, which is the youngest, appears to coincide with the Local Arm of the Milky Way. Expanding the search radius beyond 1 kpc will be necessary in the future to confirm that a similar stringing occurs along it outside of the immediate solar neighborhood. However, if it is indeed a signature of the Local Arm, then the other streams could also be considered as remnants from the older spiral arm-like structures that can no longer be traced with gas. Because they have well-defined age ranges, this may suggest that the spiral arm pattern of a galaxy undergoes evolution over a period of a couple of orbital periods, thus supporting simulations that produce transient spiral arms \citep[e.g.,][]{kawata2014,pettitt2015}.

The arrangement of the strings preferentially perpendicular to the streams is also notable, as it might suggest a major mode of the giant molecular cloud assembly. Detailed comparison to simulations will be necessary to understand the dominant mechanisms behind it.


\appendix
\section{CNN architecture}\label{sec:cnn}

\begin{lstlisting}[language=Python]
class Net(nn.Module):
    def __init__(self, input_shape=(1, 10, 250)):
        super(Net, self).__init__()
        self.conv1 = nn.Conv2d(1, 8, 3,padding=1)
        self.conv2 = nn.Conv2d(8, 8, 3,padding=1)
        self.conv3 = nn.Conv2d(8, 16, 3,padding=1)
        self.conv4 = nn.Conv2d(16,16, 3,padding=1)
        self.conv5 = nn.Conv2d(16,32, 3,padding=1)
        self.conv6 = nn.Conv2d(32,32, 3,padding=1)
        self.conv7 = nn.Conv2d(32,64, 3,padding=1)
        self.fc1 = nn.Linear(448, 512)
        self.fc2 = nn.Linear(512, 512)
        self.fc3 = nn.Linear(512, 2)

    def _forward_features(self, x):
        x = F.max_pool2d(F.relu(self.conv2(F.relu(self.conv1(x)))), 2)
        x = F.max_pool2d(F.relu(self.conv4(F.relu(self.conv3(x)))), 2)
        x = F.relu(F.max_pool2d(self.conv5(x), 2))
        x = F.relu(F.max_pool2d(self.conv6(x), 2))
        x = F.relu(F.max_pool2d(self.conv7(x), 2))
        return x

    def forward(self, x):
        x = self._forward_features(x)
        x = x.view(x.size(0), -1)
        x = F.relu(self.fc1(x))
        x = F.relu(self.fc2(x))
        x = self.fc3(x)
        return x
\end{lstlisting}

The input are formatted in a 10$\times$250 matrix, with 10 columns sorted into 9 photometric bands and the parallaxes, and the rows containing information on the individual stars, sorted by the $M_G$ flux. The network first performs a 2d convolution, breaking this input into 8 channels, passing this output through another 2d convolution, reassembling it in the same shape. Then the matrix is dowsampled using max pooling reducing dimensionality, turning $8\times10\times250$ matrix into $8\times5\times125$. Another 2d convolution of a 2d convolution is performed, breaking it into 16 channels, and after max pooling, turning it into $16\times2\times62$ matrix. Then 2d convolution followed by max pooling is performed 3 times, resulting in $32\times2\times31$, $16\times1\times15$, and $64\times1\times7$. The resulting 448 nodes are stacked, and they are fully connected to a first hidden layer with 512 nodes, followed by a second hidden layer with 512 nodes, which is then connects to the two outputs, which are the age and the extinction value for each cluster.

\software{TOPCAT \citep{topcat}, HDBSCAN \citep{hdbscan}, PyTorch \citep{pytorch}, BASE-9 \citep{base9}, GalPy \citep{galpy}}

\acknowledgments
We thank Tristan Cantat-Gaudin for the wonderful discussion. Additionally, we acknowledge Brian Hutchinson and Richard Olney in regards to the assistance in construction of the CNN, and Ted von Hippel and Elliot Robinson in helping to set up BASE-9. M.K. and K.C. acknowledge support provided by the NSF through grant AST-1449476, and from the Research Corporation via a Time Domain Astrophysics Scialog award (\#24217).
This work has made use of data from the European Space Agency (ESA)
mission {\it Gaia} (\url{https://www.cosmos.esa.int/gaia}), processed by
the {\it Gaia} Data Processing and Analysis Consortium (DPAC,
\url{https://www.cosmos.esa.int/web/gaia/dpac/consortium}). Funding
for the DPAC has been provided by national institutions, in particular
the institutions participating in the {\it Gaia} Multilateral Agreement.

\bibliographystyle{aasjournal.bst}
\bibliography{strings.bbl}

\end{document}